\documentclass[twocolumn]{aastex7}

\usepackage{longtable}
\usepackage{amsmath}
\usepackage{graphicx}
\usepackage{hyperref}
\usepackage{makecell}

\begin{document}

\title{X-rays Mark the Spot: The Effects of Reduced Metallicity on X-ray AGN Obscuration at High Redshift}

\correspondingauthor{Yash A. Gursahani}

\author[0000-0003-1754-2570]{Yash A. Gursahani}
\affiliation{Department of Astronomy, University of Maryland, College Park, 4296 Stadium Dr, College Park, MD 20742, USA}
\email[show]{yashag [at] umd [dot] edu}

\author[0000-0002-1510-4860]{Christopher S. Reynolds}
\affiliation{Department of Astronomy, University of Maryland, College Park, 4296 Stadium Dr, College Park, MD 20742, USA}
\affiliation{Joint Space-Science Institute, College Park, MD 20742, USA}
\email{creynold@umd.edu}



\begin{abstract}
The James Webb Space Telescope has pushed the frontier of high-redshift galaxy and active galactic nucleus (AGN) observations firmly past $z=10$. Corresponding to the first 500\,Myr after the Big Bang, this coincides with the epoch of supermassive black hole seeding and their early growth, much of which is likely to occur in highly obscured environments. In this work, we investigate the expected X-ray properties of these obscured AGNs focusing on the impact of the significantly lower iron abundance predicted at such early times.  We use Monte Carlo methods to model the radiative transfer of X-rays from a central AGN through a surrounding torus of cold gas, characterizing the emergent X-ray spectrum as a function of the metallicity, opening angle of the torus, and column density.  Motivated by expectations of high-$z$ systems, we focus on Compton-thick obscurers with columns $N_H=10^{24}-10^{25}\,{\rm cm}^{-2}$.  We find that decreased metallicity can significantly increase the fraction of X-ray photons that escape the torus, improving the prospects of detecting these very high-$z$ AGNs. The covering fraction of the obscurer (i.e. torus opening angle) plays a complex role, with repeated scatterings across the interior of the torus (isotropizing the emission) competing with escape through the opening, producing geometric beaming. Additionally, we explore non-solar abundance ratios that mimic the delay-time distribution of Type Ia supernovae. We use our models to address the detectability of highly obscured $z=10$ AGNs in next-generation, high-angular resolution X-ray surveys.
\end{abstract}


\section{Introduction} \label{sec:intro}

Few terms in astronomy have evolved as quickly as the ``high" in ``high-redshift." When the Hubble Space Telescope (\textit{HST}) commenced its observing program in 1993, $z\sim3-4$ was considered the frontier of high-redshift galaxy observations (e.g. \citealt{steidel_deep_1993}, \citealt{press_properties_1993}, \citealt{eales_infrared_1993}). \textit{HST} pushed this frontier to $z\sim11$ with the discovery of GN-z11 \citep{oesch_remarkably_2016}. Since the launch of the James Webb Space Telescope (\textit{JWST}) in 2021, the high-redshift landscape has shifted once again. The record-holder at the time of writing for the most distant, spectroscopically confirmed galaxy is MoM-z14 at a redshift of 14.44 \citep{naidu_cosmic_2026}. Such an extreme redshift corresponds to only 283 Myr after the Big Bang. \textit{JWST} continues to increase the sample of known galaxies at $z\gtrsim10$ at an unprecedented rate.

There is often uncertainty, in particular for galaxies showing strong optical emission lines, whether the source of this emission is predominantly star formation in star-forming galaxies (SFGs) or an active galactic nucleus (AGN). Several diagnostic tools have been developed to disentangle various members of this ``galactic zoo," perhaps the most prominent being the Baldwin, Phillips \& Terlevich (BPT) diagram \citep{baldwin_classification_1981}. Other diagnostics have been suggested by, for example, \citet{veilleux_spectral_1987}, \citet{kewley_optical_2001}, \citet{stasinska_semi-empirical_2006}, \citet{juneau_new_2011}, \citet{yan_aegis_2011}, and \citet{feuillet_classifying_2024}, to name a few. At high redshift, \citet{mazzolari_new_2024} have tested diagnostics based on the [OIII]$\lambda$4363 auroral line. However, many sources at $z \gtrsim 5$ are still not uniquely identified as AGNs or SFGs. For example, \citet{mazzolari_new_2024} place GN-z11 in the ``SFG or AGN" portion of the diagrams.

Though a broad range of astrophysical objects can produce X-ray emission within a galaxy, an X-ray luminosity $> 10^{42}$ erg s$^{-1}$ from a nuclear point source is thought to be a telltale signature of an accreting, supermassive black hole (e.g. \citealt{bowyer_detection_1970}, \citealt{elvis_seyfert_1978}). A number of \textit{JWST} targets show some indication of accretion by supermassive black holes (SMBHs) that power AGNs. UHZ1 \citep{bogdan_evidence_2024} and GHZ9 (\citealt{kovacs_candidate_2024}, \citealt{napolitano_dual_2025}) have been claimed to host AGNs. Both at $z\sim10$, these galaxies are lensed by the foreground cluster Abell 2744 and were observed by \textit{JWST}. The AGN classification follows from the co-spatial detection of X-rays in deep imaging from the \textit{Chandra} observatory \footnote{We note that the recent re-analysis by \citet{zou_revisiting_2026} suggests that the significance of this X-ray detection could be lower than previously claimed.}. UHZ1 has quite a poorly-constrained column density, though it could be as high as $8 \times 10^{24}$ cm$^{-2}$, requiring an extremely luminous AGN in the 2-10 keV band with $L_X\sim10^{46}$ erg s$^{-1}$ \citep{bogdan_evidence_2024}. In the local Universe, Compton-thick AGNs (CT AGNs), or those with log($N_H$/cm$^{-2}$) $\geq 24$, are difficult to observe due to the large absorption opacity and Compton scattering. \citet{ricci_compton-thick_2015} found that while only $\sim8\%$ of these sources are detected in the 70-month \textit{Swift}/BAT catalog, the intrinsic fraction of CT AGNs is roughly 27\%. \citet{georgantopoulos_compton-thick_2025} find a similar value of $\sim24\%$, while other studies find even higher intrinsic fractions in the BAT-selected sample (\citealt{torres-alba_compton-thick_2021}, \citealt{tanimoto_nustar_2022}). These results are corroborated by so-called ``isotropic" selection methods in the optical, infrared, and radio (\citealt{boorman_nustar_2024}, \citealt{annuar_compton-thick_2025}). Isotropic methods aim to classify CT AGNs without relying on flux-limited, hard X-ray selection which often suffers from model-dependent corrections. Taken together, there is a wealth of evidence showing a clear observational bias against finding CT AGNs at low redshift. 

This motivates an investigation into the X-ray spectral properties of obscured AGNs at high redshift. If these objects are indeed observable, their discovery would place them among the earliest known black holes in the Universe. Their characterization would also have sizable implications for studies of black hole seeding, their growth over cosmic time, and their coevolution with their host galaxies.

The origin of X-ray emission in AGNs is still debated, but is generally thought to result from the Comptonization of thermal accretion disk photons by hot electrons in a corona (e.g. \citealt{bisnovatyi-kogan_disk_1977}, \citealt{haardt_two-phase_1991} \citealt{nandra_ginga_1994}). This upscattering produces a power-law spectrum which we take as the source spectrum in this work, remaining agnostic about coronal properties and geometry. AGNs are thought to be surrounded by a dusty, neutral absorber, perhaps arranged in a toroidal geometry and located at $\sim10^5-10^6 $ gravitational radii ($\sim$pc scales) from the black hole \citep{hickox_obscured_2018}. In this obscuring material, X-rays are primarily lost to photoelectric absorption and subsequent Auger electron ejection. Elements such as carbon, oxygen, magnesium, and others readily absorb X-rays between $\sim 0.1 - 10$ keV. In particular, the solar abundance of iron is on the order of $10^{-5}$ relative to hydrogen but the K-shell cross section is about a factor of $10^{5}$ larger than the Thomson cross section. This means that in the local Universe, CT obscuration coincides with the iron K-shell being optically thick to X-rays with energies between about 7 and 10 keV. Cold electrons bound to hydrogen or helium atoms also Compton scatter photons, redirecting them and reducing their energy. In obscurers with $\tau_e > 1$, photons scatter $\sim \tau_e^2$ times, where $\tau_e = N_H \sigma_T$ is the electron scattering optical depth. This increases the effective path length by a factor of $\tau_e$ and therefore the probability of absorption by metals, naturally explaining the lack of observed CT AGNs nearby. 

However, accreting black holes at high redshift may not be surrounded by solar-metallicity gas. The chemical enrichment of the Universe is a gradual process, relying on the details of stellar evolution and the baryon cycle (e.g. \citealt{kobayashi_history_2000}, \citealt{kobayashi_origin_2020}). Therefore, one might expect that at $z\gtrsim10$, galaxies may not have had time for enough generations of massive stars to form, detonate as supernovae, and disperse an appreciable amount of heavy elements across the interstellar and circumgalactic media (ISM and CGM, respectively). Type Ia supernovae (SNe Ia), which provide the majority of the iron in the Universe, result from low-mass stars and hence have an additional delay-time distribution (DTD; e.g. \citealt{mannucci_two_2006}, \citealt{ruiter_rates_2009}, \citealt{mennekens_delay-time_2010}, \citealt{strolger_delay_2020}). \citet{ruiter_rates_2009} find that the double-degenerate white dwarf scenario reproduces the observed power-law SNe Ia rate through cosmic time. Though there may be a few SNe Ia that explode within 100 Myr of the stellar population forming, the majority are thought to occur over the course of a Hubble time. Since they require low-mass stellar remnants, the onset of Type Ia detonations would not be coincident with the first generation of massive stars. A consequence of this is lower metallicity and perhaps non-solar abundance ratios in the early Universe. Low abundances of iron and other elements at high redshift may allow more X-ray photons to pass through undeterred to the observer.

Another consideration is that both the electron cross section and Compton scattering angle depend on photon energy. For harder photons, the angle-integrated Klein-Nishina cross section \citep{klein_uber_1929} decreases by a modest amount compared to the Thomson cross section. This slightly reduces their chance of scattering compared to softer photons. If scattering does occur, the likelihood for a photon with $E \gtrsim 10$ keV to be scattered forward is higher than for lower energies. That is, more hard X-rays will on average continue in their original direction rather than scattering in a random direction. We might then expect that the rest-frame, hard X-rays scatter fewer than the nominal $\tau_e^2$ times. They therefore have a better chance of making it out of the obscuring material.

Monte Carlo calculations have been an integral tool in studying the interactions of X-rays with gas surrounding AGNs since the work of \citet{lightman_effects_1988}. These techniques were furthered by, e.g. \citet{george_x-ray_1991}, \citet{reynolds_reflection-dominated_1994}, \citet{magdziarz_angle-dependent_1995}, \citet{nandra_xmm-newton_2007}, and \citet{ikeda_study_2009}. \citet{murphy_x-ray_2009} developed the \textsc{mytorus} model, and we reference this paper often in our work as MY09. In recent years, several groups (e.g. \citealt{brightman_xmmnewton_2011}, \citealt{paltani_reflex_2017}, \citealt{balokovic_new_2018}, \citealt{buchner_x-ray_2019}, \citealt{tanimoto_xclumpy_2019}, \citealt{meulen_x-ray_2023}, \citealt{ricci_ray-tracing_2023}) have extended these types of models, making them more flexible by adding additional free parameters and geometries. Notably, most of these allow for variable elemental abundances, and the model presented in \citet{ricci_ray-tracing_2023} includes effects from dust as well.

We continue this tradition, employing Monte Carlo methods in our modeling of high-redshift AGNs. In this paper, we examine the effects of low-metallicity environments around these objects, such as may be found in the early Universe. We find that reduced elemental abundances result in a lower flux suppression with redshift. Section \ref{sec:methods} will outline the details of our simulation framework. In Section \ref{sec:results}, we show the results of our calculations for a subset of the parameter space. Section \ref{sec:discussion} contains a discussion of our results and their observational implications. Finally, we summarize our work and state our conclusions in Section \ref{sec:conclusion}. Throughout this work, we assume a $\Lambda$CDM cosmology with $H_0=70$ km s$^{-1}$ Mpc$^{-1}$, $\Omega_m=0.27$, and $\Omega_{\Lambda}=0.73$.

\section{Methods} \label{sec:methods}
We implement a 3-dimensional Monte Carlo code to follow photons from an intrinsic X-ray source through cold matter arranged in a toroidal geometry around the BH and accretion disk. In order to carry out such calculations, we use the Mersenne Twister algorithm to generate pseudo-random numbers \citep{matsumoto_mersenne_1998}. Each simulation tracks $1.5 \times 10^8$ photons, which are emitted isotropically from the central X-ray source. The photon energies are chosen from a uniform probability distribution in logarithmic space according to 

\begin{equation} \label{eq:log_uniform}
    P(E) = \frac{1}{E \text{ log}(E_{\text{max}}/E_{\text{min}})}
\end{equation}

\noindent where $E_{\text{min}}$ = 0.2 keV and $E_{\text{max}}$ = 500 keV. Since there are far more photons at soft energies, this method of initialization more densely samples lower energies and therefore reduces noise in this regime. As the photons undergo reprocessing by atoms, we record their initial energies, final energies, number of Compton scatterings, and final direction. This allows us to weight the output according to any input photon spectrum at the source, which we expect will take the form of a power-law of index $\Gamma$ from the BH corona: $N_{\text{phot}}(E) \propto E^{-\Gamma}$. It is possible that the finite coronal temperature imposes a high-energy ($\gtrsim 100$ keV) cutoff in the source spectrum, which could affect the transmitted flux at high redshift. For clarity, we do not include such a cutoff in our power-law, though this could be easily implemented. Table \ref{tab:param_space} describes the range of parameters over which we conducted our runs.

We note that this is equivalent to the Green's function approach used by MY09 and others. In this framework, one would fire a beam of photons with a single energy into a medium with some column density, geometry, and metallicity. For some final photon direction, the Green's function $G(E_i, E_f, N_H, g_1,...,g_K, Z_1,...,Z_M)$ is the spectrum resulting from the mono-energetic beam's interactions with the gas. Here, $N_H$ is the column density, $g_1,...,g_K$ are the $K$ parameters describing the geometry, and $Z_1,...,Z_M$ are the abundances of $M$ species in the gas. This process is repeated for beams with a range of energies. To obtain an output spectrum, we must also consider the shape of the input spectrum (a power-law in our case). We denote the input spectrum by $I(E)$. Then, the output spectrum $O(E)$ is the integral of the individual Green's functions over initial energy, weighted by the input. That is, 

\begin{equation} \label{eq:greens}
\begin{split}
    O(E) = \int_{E_{\text{min}}}^{E_{\text{max}}} &G(E', E, N_H, g_1,...g_K, Z_1,...Z_M) \\ 
    &\times I(E') \text{~d}E'.
\end{split}
\end{equation}

\noindent The Monte Carlo simulation gives us a reasonable approximation to the integral by discretizing the initial photon energies. We emphasize here that this is a linear problem; our output spectra could be decomposed into Green's functions even though we do not directly use this approach.

\begin{table}[h]
    \centering
    \begin{tabular}{c|c|c}
    Parameter & Range & Step \\
    \hline
    \hline
    log ($N_H \text{/cm}^{-2}$) &  21.0 - 25.0 & 0.5 \\
    \hline
    Opening Angle & 30$^{\circ}$ - 120$^{\circ}$ & 30$^{\circ}$ \\
    \hline
    Metallicity & \makecell{0, 0.01$Z_{\odot}$, 0.1$Z_{\odot}$, 1$Z_{\odot}$, \\ NS$_5$, NS$_{10}$, NS$_{50}$} & --
    \end{tabular}
    \caption{Parameter space spanned by our Monte Carlo simulations. Multiples of $Z_{\odot}$ refer to metallicities in which all 11 elements are in solar ratios. Non-solar (`NS$_X$') metallicity prescriptions indicate that iron-peak elements (Cr, Fe, and Ni) are held fixed at 1\% of their solar value which the remaining 8 elements are tuned to $X\%$ of their solar value. This implements non-solar abundance ratios to reflect the diminished importance of Type Ia supernovae at early times.}
    \label{tab:param_space}
\end{table}

\subsection{Geometry} \label{subsec:geometry}
\begin{figure}[h]
    \centering
    %\hspace{-3cm}
    \includegraphics[width=\linewidth]{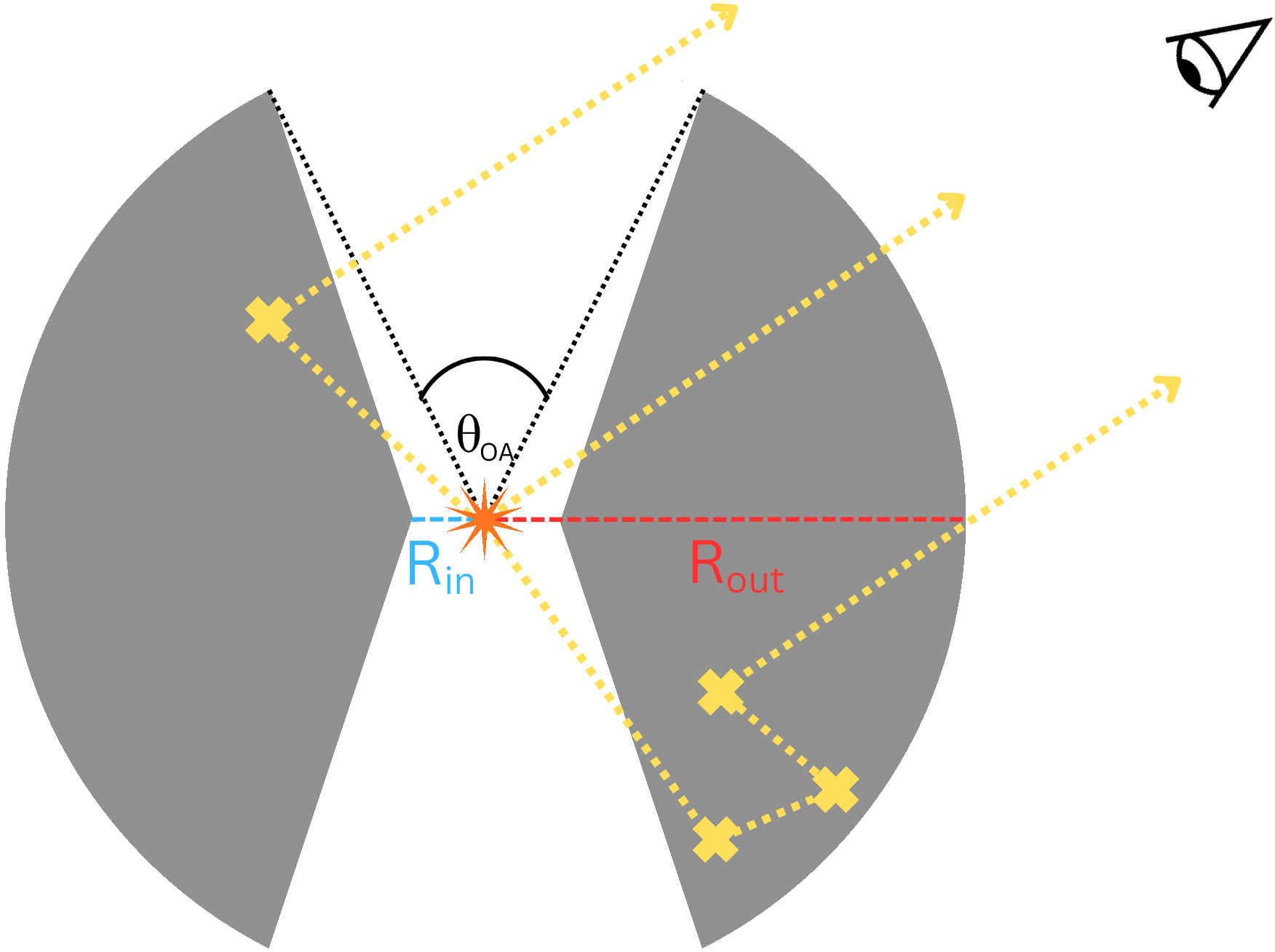}
    \caption{The geometry used in our Monte Carlo code. We use $Y = R_{\text{out}}/R_{\text{in}} = 50$ and a range of opening angles $\theta_{\text{OA}} = 30^{\circ}$, $60^{\circ}$, $90^{\circ}$, $120^{\circ}$. The X-ray source (orange star) is placed at the origin. Shown here is one example of a viewing angle, where all photons traced in yellow will reach the observer after undergoing interactions (yellow `x' marks) with the torus.}
    \label{fig:geometry}
\end{figure}

We use a spherical-toroidal geometry, similar to that in \citet{ikeda_study_2009}. A cross section through the torus is shown in Figure \ref{fig:geometry}. We stress that we do not resolve the accretion disk or any of its physics, including reprocessing by ions in the plasma, relativistic blurring of spectral features by the inner disk, or warping of photon trajectories in the strong gravitational well of the BH. Instead, we assume a power-law of photon energies which result from inverse Compton scattering of thermal disk photons by a non-thermal electron population in the central corona. The geometry, physical extent, and precise location of the corona is debated, but the power-law at hard energies has been observed for several decades in AGNs (e.g. \citealt{elvis_atlas_1994}, \citealt{nandra_ginga_1994}). Though these coronal properties are all subjects of ongoing work, it is generally accepted that the corona resides within a few gravitational radii of the BH itself. Since the scales we are concerned with here are far larger than this, we place our X-ray source at the origin for simplicity. 

% We choose an inner radius of 0.5 pc in accordance with \citet{li_constraining_2023}, who model the torus given reverberation lag between optical photometry from SDSS and mid-infrared dust emission in WISE band W1. Our outer radius is 0.75 pc, so for photons emitted along edge-on ($i = 90^{\circ}$) lines of sight, the shortest distance they travel through the torus is $\ell = 0.25$ pc.

The cold medium that surrounds the source can be thought of as the ``dusty torus" inferred from optical and infrared observations of AGNs.  The medium with which photons can interact is uniformly distributed, characterized only by the hydrogen column density $N_{\text{H}} = n_{H} \ell$, where $\ell = R_\text{out} - R_{\text{in}}$ is the distance through the toroidal equator. For any value of $N_{H}$ with which we wish to run our simulation, we simply invert this relationship to obtain $n_{H}$. Our choice of units for $R_{\text{in}}$ and $R_{\text{out}}$ is therefore unimportant, though the dimensionless ratio $Y = R_{\text{out}}/R_{\text{in}}$ does affect the geometry. We run all our calculations with $Y=50$, and \citet{ikeda_study_2009} show that the resulting spectra are only weakly dependent on this parameter in the range $10 < Y < 100$. A smaller value, such as $Y\sim1$, could make the toroidal walls concave rather than the convex shape we show in \ref{fig:geometry} and significantly affect the results. After each step the photon is propagated, there are a number of possible interactions it can have with the medium, which are described in the next section. When the photon escapes or is absorbed, we record the photon's final energy, position, direction, and number of scattering events. We explain the escape conditions in detail in Appendix \ref{subsec:app_geo}.

% Future work that may include a density gradient or a ``clumpy" torus would be more sensitive to the choice of distance scales.

% \textbf{We fix our step size such that a photon traveling unhindered through the equator would take $3\times10^3$ steps to escape.} For a CT medium with $N_{H} = 1/\sigma_T = 1.5\times10^{24}$ $\text{cm}^{-2}$ (where $\sigma_T$ is the Thomson cross section), this yields an \textbf{optical depth per step of $\Delta\tau = 3.33 \times 10^{-4}$}. 

After all photons have escaped or been absorbed, we use their final direction to assign them to one of ten bins of equal solid angle. These represent ten possible viewing angles for the system and are the same bins as those described by MY09. Each bin covers a range of inclinations $i$, which we reproduce in Table \ref{tab:sa_bins}. As in MY09, we choose this binning system so that all bins receive an identical number of input photons from the source, namely $1.5 \times 10^7$ of them. Depending on the geometry, column depth, opening angle, and metallicity, the eventual fate of these photons will vary. We discuss this further in Section \ref{subsec:spectra}.

\begin{table}[h]
    \centering
    \begin{tabular}{c|c|c|c|c}
    Bin & cos $i_{\text{max}}$ & cos $i_{\text{min}}$ & $i_{\text{min }} (^\circ)$ & $i_{\text{max }} (^\circ)$ \\
    \hline
    1 & 0.9 & 1.0 & 0.00 & 25.84 \\
    2 & 0.8 & 0.9 & 25.84 & 36.87 \\
    3 & 0.7 & 0.8 & 36.87 & 45.57 \\
    4 & 0.6 & 0.7 & 45.57 & 53.13 \\
    5 & 0.5 & 0.6 & 53.13 & 60.00 \\
    6 & 0.4 & 0.5 & 60.00 & 66.42 \\
    7 & 0.3 & 0.4 & 66.42 & 72.54 \\
    8 & 0.2 & 0.3 & 72.54 & 78.46 \\
    9 & 0.1 & 0.2 & 78.46 & 84.26 \\
    10 & 0.0 & 0.1 & 84.26 & 90.00 \\
    \hline
    \end{tabular}
    \caption{Our solid angle binning scheme. Note that the bins are equally spaced in cos $i$, but not in $i$ itself. Values are taken from Table 1 in MY09.}
    \label{tab:sa_bins}
\end{table}

\subsection{Photon Interactions} \label{subsec:interactions}
We implement two possible interactions between photons and matter in our simulation: Compton scattering and photoelectric absorption by metals, which could result in either fluorescent line emission or the Auger effect. We stress here that since we assume a cold medium, we do not consider the effect of ionized species such as those expected in the accretion disk and broad line region. This is justified since we are interested in the effects of the dusty torus at much larger physical scales.

In order to determine whether an interaction occurs at each step, we first calculate the differential optical depth along the step: $\Delta\tau = n\sigma_{\text{tot}}ds$, where 

\begin{equation} \label{eq:sigma_tot}
\begin{split}
    \sigma_{\text{tot}}(E) &= \sigma_{\text{scat}}(E) + \sigma_{\text{abs}}(E) \\
    &= \sigma_{\text{scat}}(E) + \sum_{i=1}^N Z_i \sigma_i(E).
\end{split}
\end{equation}

\noindent Here, $\sigma_{\text{scat}}$ denotes the cross section for Compton scattering while $\sigma_i$ is the photoelectric cross section for each of $N$ atoms other than hydrogen, from which either a fluorescent line photon will emerge or to which the incident photon is lost to the Auger effect. $Z_i$ is the abundance of the element relative to hydrogen; this makes it straightforward to implement non-solar abundance ratios. Solar photospheric abundance values are taken from \citet{lodders_solar_2025} and the metallicity schemes we use are shown in Table \ref{tab:param_space}.

The interaction probability is then given by the usual exponential attenuation formula

\begin{equation} \label{eq:p_int}
    P_{int} = 1 - e^{-\Delta\tau}
\end{equation}

\noindent and we draw a uniform random variable in order to decide whether the photon continues to the next step or undergoes one of the three aforementioned processes. If an interaction occurs, another random variable draw determines whether the photon undergoes a scattering event or an atomic interaction with a particular element. The following two sections will expand on our treatment of Compton scattering and the atomic physics, respectively.

\subsubsection{Compton Scattering} \label{subsubsec:compton}
In our code, photons can scatter off electrons bound to hydrogen or helium. Following MY09, we multiply the scattering optical depth from hydrogen by a factor of 11/9 to account for the electrons contained in helium. For all photon energies, we use the full Klein-Nishina cross section for $\sigma_\text{scat}$. With $x \equiv E_{\text{phot}}/m_{e}c^2$, this is 

\begin{equation}
\begin{split}
    \sigma_{\text{KN}} &= \sigma_\text{T} \times \frac{3}{4} \left\{ \frac{1+x}{x^3} \left[ \frac{2x(1+x)}{1+2x} - \text{ln}(1+2x) \right] \right. \\
    &\left. + \frac{1}{2x}\text{ln}(1+2x) - \frac{1+3x}{(1+2x)^2} \right\}.
\end{split}
\end{equation}

\noindent where $\sigma_\text{T} = 6.65 \times 10^{-25}$ cm$^{2}$ is the Thomson cross section.

Now, we must redirect the scattered photon. Since we do not incorporate polarization in our scattering algorithm, the azimuthal angle $\phi$ is chosen uniformly. Selecting the polar angle $\theta$ requires more care. The Klein-Nishina differential cross section has a non-trivial dependence on $\theta$ as well as the photon energy; scattering is forward-peaked at higher energies. In our Monte Carlo approach, we choose $\theta$ probabilistically with a random number. First, let $E$ and $E'$ denote our pre- and post-scattering energies, respectively. The differential cross section is 

\begin{equation}
    \frac{\text{d}\sigma}{\text{d}\Omega} = \frac{r_0^2}{2}\frac{E'^2}{E^2} \left( \frac{E}{E'} + \frac{E'}{E} - \text{sin}^2\theta \right)
\end{equation}

\noindent where $r_0 = 2.818 \times 10^{-13}$ cm is the classical electron radius and $E'$ is given by the Compton formula (see, e.g. \citealt{rybicki_radiative_1979}):

\begin{equation} \label{eq:compton}
    E' = \frac{E}{1 + \frac{E}{mc^2}(1 - \text{cos}~\theta)}
\end{equation}

\noindent Since $\text{d}\Omega = \text{sin} \, \theta \, \text{d}\theta \, \text{d}\phi$, we define

\begin{equation} 
\begin{split}
    \sigma(\theta) \equiv \int \frac{\text{d}\sigma}{\text{d}\theta'} \text{d}\theta'
    = \int^{2\pi}_{0} \int^{\theta}_{0} \frac{\text{d}\sigma}{\text{d}\Omega} \text{sin} \, \theta' \text{d}\theta' \text{d}\phi. 
\end{split}
\end{equation}

\noindent This integral does have an analytic solution, though we omit it here for brevity. Let $\eta \equiv \sigma(\theta)/\sigma_{\text{scat}} \in [0, 1]$ be our random number, drawn uniformly. Our goal is to invert $\sigma(\theta)$ and arrive at the correct scattering angle. 

To find the angle $\theta$ which corresponds to our random number $\eta$ and energy $E$, we use Newton-Raphson root finding to solve the equation $\eta - \frac{\sigma(\theta)}{\sigma_{\text{scat}}} = 0$ for $\theta$. This is done on a grid of 5000 photon energies and 1000 values of $\eta \in [0, 1]$. Similarly to photon initialization, we space the energies logarithmically. This produces a lookup table of angles for each pre-scattering energy and random number in the grid. 

\subsubsection{Photoelectric Absorption and Line Emission} 
\label{subsubsec:phabs_line}

The atomic processes we treat here are K-shell photoelectric absorption and fluorescent line emission from 11 elements, listed here in order of atomic number: C, O, Ne, Mg, Si, S, Ar, Ca, Cr, Fe, and Ni. Due to its significant optical depth at high column density, we also include L-shell absorption and fluorescence from Fe.

To calculate the $\sigma_{\text{abs}}$ term in Equation \ref{eq:sigma_tot} for each element, we use the fits from \citet{verner_analytic_1995}. The formula given here is $\sigma_{\text{abs}}(E) = \sigma_0 F(y)$, where 

\begin{equation} \label{eq:verner}
    F(y) = y^{(0.5P - 5.5)} [(y - 1)^2 + y_w^2] (1 + \sqrt{y/y_a})^{-P}.
\end{equation}

\noindent Here, $y = E/E_0$ while $\sigma_0$, $E_0$, $y_w$, $y_a$, and $P$ are fit parameters. An additional parameter to consider is $E_{\text{th}}$, the threshold energy below which absorption cannot occur so that we set $\sigma_{\text{abs}}(E) = 0$. At the initialization of each photon and at each subsequent change in its energy, K-shell (1s) cross sections for all 11 elements are calculated and stored. We calculate cross sections for the 2s and 2p shells of iron separately, since only absorption by the 2p shell can result in fluorescence.

When a particular element is selected to undergo an interaction with a photon, we must decide whether the absorption process will result in Auger electron ejection or line emission. This is probabilistic in nature, with the chance of line emission given by the fluorescent yield of a species. The remaining probability is assigned to the Auger effect, and we draw a random number to decide between the two. We use yields from \citet{meddouh_average_2023}. 

If line emission occurs, we set the energy of the photon to the respective K$\alpha$ line for most elements. For iron, we include more complexity. For Fe K-shell fluorescence, there is a $\sim10\%$ probability that excitation produces a Fe K$\beta$ line at 7.06 keV. We also split the Fe K$\alpha$ line into Fe K$\alpha_1$ at 6.404 keV and Fe K$\alpha_2$ at 6.391 keV according to their branching ratio of 2:1. For Fe L-shell fluorescence, there is a $\sim63\%$ chance that L$\alpha$ (0.705 keV) photons are produced, while the rest are L$\beta$ photons (0.719 keV). All line energies and branching ratios are taken from \citet{thompson_x-ray_2009}. Fluorescent photons have a randomly selected direction in both $\theta$ and $\phi$, since atoms emit them isotropically. 

If instead Auger ejection is the selected process, we set the final photon energy to zero and terminate its path, as the photon is now lost. Keeping count of how many photons are absorbed will be important for later metrics, such as the transmitted fraction of photons in some energy band. 

To summarize this section, we determine if the photon interacts with the medium via Equation \ref{eq:p_int}. If this interaction is Compton scattering, we then choose a new azimuthal angle $\phi$ isotropically. We draw a random number $\eta$ between 0 and 1, convert the photon energy to a discrete index, and use our lookup table to choose $\theta$. Finally, with $\theta$, we can use Equation \ref{eq:compton} to find post-scattering energy of the photon. If the interaction is atomic, we choose the element that will absorb the photon. Then, we decide whether the atom undergoes fluorescent photon emission or Auger ejection via the fluorescent yield. Any line photons are emitted in a random direction. 

\subsection{Equivalent Width and Compton Shoulder} \label{subsec:eqw}

The primary diagnostic we use to determine the strength of a line is its equivalent width (EW), defined as 

\begin{equation}
    \text{EW} \equiv \int_{E_\text{low}}^{E_\text{high}} \frac{F_s(E) - F_c(E)}{F_c(E)} dE
\end{equation}

\noindent where $E_\text{low}$ and $E_\text{high}$ are limits set by the width of the line in units of energy, $F_s$ is the spectrum including the line, and $F_c$ is the local continuum without the line. We use the \texttt{quad} integration method from the \texttt{scipy.integrate} package and the \texttt{Akima1DInterpolator} within the \texttt{scipy.interpolate} package to represent $F_s$ as a continuous function of energy.

While this is a relatively straightforward quantity to calculate, X-ray spectra pose a challenge. Line photons, which are emitted at one characteristic energy in the rest frame of the atom, can subsequently undergo one or several Compton scatterings. In a CT, cold medium such as that implemented in this work, post-scattering photon energies are always reduced. Because of the excess of photons in the line, it develops a ``shoulder" of down-scattered photons called the Compton shoulder (CS). The width of the first-order CS is approximately two Compton wavelengths, defined as $\lambda_{\text{c}} = h/m_e c = 0.02426$ $\text{\AA}$. Doubling this corresponds to $\Delta E_{\text{CS}} \approx 160$ eV for the Fe K$\alpha_2$ line, as an example. Of course, photons in the first-order CS can scatter down to lower energies, producing a second-order CS, and so on. In this work, we only focus on the first-order feature, which appears clearly in our simulated spectra.

The CS of the Fe K$\alpha$ line has only been resolved in a handful of observations, owing to the physical complexity of the Fe K complex and instrument resolution (e.g. \citealt{kaspi_ionized_2002}, \citealt{bianchi_flux_2002}, \citealt{watanabe_detection_2003}, \citealt{matt_high_2004}). Doppler motion, different ionization states, and the line-of-sight column depth all play a role in shaping the spectral line. Additionally, we note here that a harder input spectrum produces a more pronounced CS \citep{odaka_sensitivity_2016} since more of the iron fluorescence occurs deeper within the layer of gas. Observations with \textit{XRISM}/Resolve have recently detected the Fe K$\alpha$ CS in the Circinus galaxy with a spectral resolution of less than 5 eV at 6.4 keV \citep{xrism_collaboration_accurate_2026}.

Though we may soon obtain a larger sample of CS measurements from AGNs, we assume in this work that the CS is blended with the line for purposes of calculating its EW. With this choice, we must model the continuum $F_c$ to perform an accurate subtraction around the line \textit{and} its shoulder. Following \citet{odaka_sensitivity_2016}, we use a model of the form

\begin{equation} \label{eq:continuum_fit}
    F_c(E) = Ax^{C + Bx}
\end{equation}

\noindent where $x \equiv E / E_{\text{line}}$ and $A$, $B$, and $C$ are free parameters. Fits are carried out using \texttt{curve\_fit} from the \texttt{scipy.optimize} library. Since we only require that the model apply to a given line, we must choose an appropriate range of values to fit with Equation \ref{eq:continuum_fit}. The energy range should be sufficiently wide for capturing the local continuum curvature, but narrow enough to avoid other spectral lines or absorption edges. For most lines, we use the spectrum 0.2 keV below the CS and 0.2 keV above the line for the fit.

Two lines that, perhaps unsurprisingly, must be treated separately from the others in our simulations are the two K-shell lines we include from iron. The Fe K$\alpha$ line proves challenging due to its prominent shoulder; an accurate model for $F_c$ is obtained by fitting the continuum over a larger portion of the spectrum, which we choose to be 0.5 keV below the CS and above the line instead of 0.2 keV. The Fe K$\beta$ line has an energy of 7.06 keV, while the Fe K edge is at 7.11 keV. We cannot use the nominal 0.2 keV range here either, as the edge introduces a sharp discontinuity. Our compromise is to extend the fit range to 0.5 keV below the line and 0.05 keV above it, just before the threshold energy for the edge. 

Finally, we note that EWs are physically meaningless without a continuum. If the continuum has zero photons in a particular energy range, the lines there have infinite EWs. This occurs for some portion of our parameter space. For edge-on viewing angles at even moderate column densities ($N_H \gtrsim 10^{22.5}$ cm$^{-2}$), photons with $E \lesssim 1$ keV are effectively absorbed by carbon and oxygen. Lines may still be present, but the continuum disappears completely. In these cases, the measured EWs should be disregarded.

\begin{figure}[h]
\hspace{-1cm}
    \centering
    \includegraphics[width=0.9\linewidth]{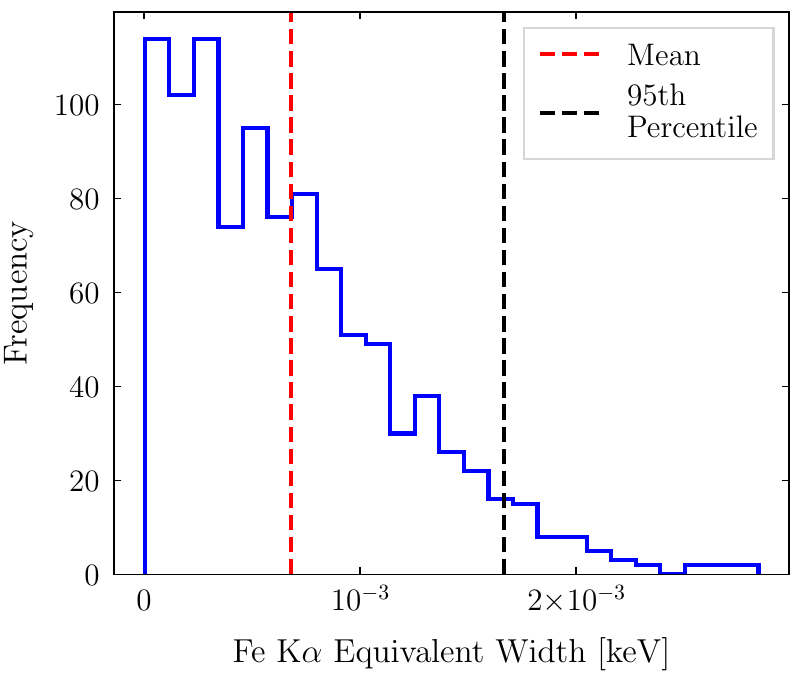}
    \caption{Histogram of Fe K$\alpha$ equivalent widths measured from 1000 realizations of a continuum-only spectrum. These values are derived only from the noise inherent to Monte Carlo energy initialization and do not represent actual line emission. The mean is located at $6.79 \times 10^{-4}$ keV and the 95th percentile is $1.66 \times 10^{-3}$ keV.}
    \label{fig:eqw_noise}
\end{figure}

As we use Monte Carlo methods in this work, we must contend with Monte Carlo noise. We would like to understand, for example, whether our calculated EW comes from the presence of a line or some insignificant fluctuations around the true power-law distribution. We quantify this in the following way. First, we draw $1.5\times10^7$ photons according to the distribution in Equation \ref{eq:log_uniform} to emulate the initial energies of photons in one solid angle bin. Then, we measure the EW around the position of a line in the same way as we describe above. Without including any physics, this measurement results from the noise alone. Repeating this process 1000 times, we can then determine the distribution of EWs from Monte Carlo noise. Figure \ref{fig:eqw_noise} shows the results for the Fe K$\alpha$ line: we obtain a 95th percentile EW of $1.66 \times 10^{-3}$ keV. We conservatively adopt this value as our lower threshold for a statistically significant EW.

\subsection{Simulated Observations} \label{subsec:sim_obs}
We make use of the \texttt{fakeit} command in \textsc{xspec} \citep{arnaud_xspec_1996} to simulate spectra from \textit{Chandra}/Advanced CCD Imaging Spectrometer-I (ACIS-I; \citealt{weisskopf_chandra_2000}, \citealt{garmire_advanced_2003}) and the Advanced X-ray Imaging Satellite (\textit{AXIS}; \citealt{reynolds_overview_2023}). \texttt{fakeit} requires a spectral model as input, so we must first convert the results of our Monte Carlo simulations to this format. We use the \textsc{HEASoft} \texttt{ftflx2tab} task to convert our simulated data to an additive table model for use in \textsc{xspec}. Each model describes a fixed opening angle, viewing angle, column density, and power-law index. The free parameters in the table model are the normalization and the redshift of the source. 

\texttt{fakeit} then allows us to set a source at a chosen redshift $z$ and with luminosity $L_X$ in some band; we use the rest-frame 2-10 keV band to normalize our model to a particular $L_X$. As for the observatories we simulate spectra from, the relevant files are the photon redistribution matrix file (RMF) and ancillary response file (ARF). Since we would like to compare next-generation missions to the \textit{Chandra} Deep Field South (CDFS), we create \textit{Chandra}/ACIS-I response files by using the \texttt{specextract} task within \textsc{ciao} \citep{fruscione_ciao_2006} on one CDFS exposure from 2007 (OBSID 08594). In reality, the CDFS is a result of combining several exposures over more than a decade of observations beginning early in the \textit{Chandra} mission. However, we choose this epoch as a midpoint which represents ACIS's capabilities before the onset of serious contamination and degradation of the soft X-ray response. \textit{AXIS} was a Probe Explorer mission concept considered by NASA for Phase A study in 2025. Although it was not selected to launch, we consider its capabilities representative of a future Probe-class or small flagship mission with arcsecond spatial resolution from $0.2-10$ keV. The response files we use for \textit{AXIS} \footnote{\url{https://axis.umd.edu/researchers/simulation-resources}} are an estimation of the true detector capabilities. We also include the non-X-ray background (NXB) and Galactic soft X-ray background models (\citealt{mccammon_high_2002}, \citealt{henley_xmm-newton_2013}, \citealt{bluem_widespread_2022}) provided by the \textit{AXIS} team for an L2 halo orbit, assuming a circular source extraction region of radius 3".

\begin{figure*}[t!]
    \centering
    \includegraphics[width=1.0\linewidth]{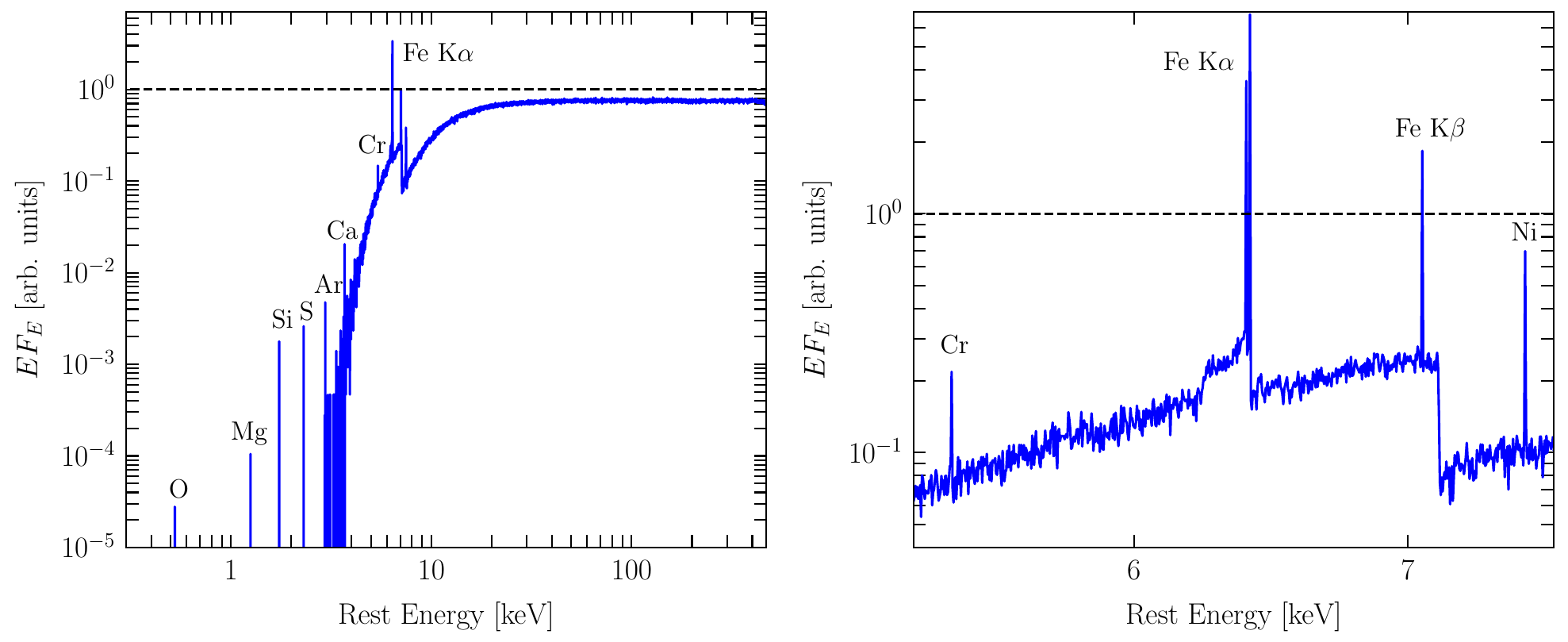}
    \caption{\textit{Left:} Example of a spectrum from our Monte Carlo simulations. This particular spectrum shows the edge-on viewing angle (Bin 10) for a geometry with $\theta_{\text{OA}} = 120^{\circ}$, solar metallicity, and $N_H = 10^{24}$ cm$^{-2}$. The black dashed line shows the $\Gamma = 2$ input spectrum. Emission lines from all labeled elements are K$\alpha$ lines unless otherwise specified. \textit{Right:} A zoomed-in version of the left panel around the Fe K$\alpha$ line and Fe K edge. The Fe K$\beta$ and Ni K$\alpha$ lines are labeled here.}
    \label{fig:basic_spectrum}
\end{figure*}

\section{Results} \label{sec:results}

\subsection{Spectra and Their Properties} \label{subsec:spectra}
\begin{figure*}[h!]
    \centering
    \includegraphics[width=\textwidth,height=\textheight,keepaspectratio]{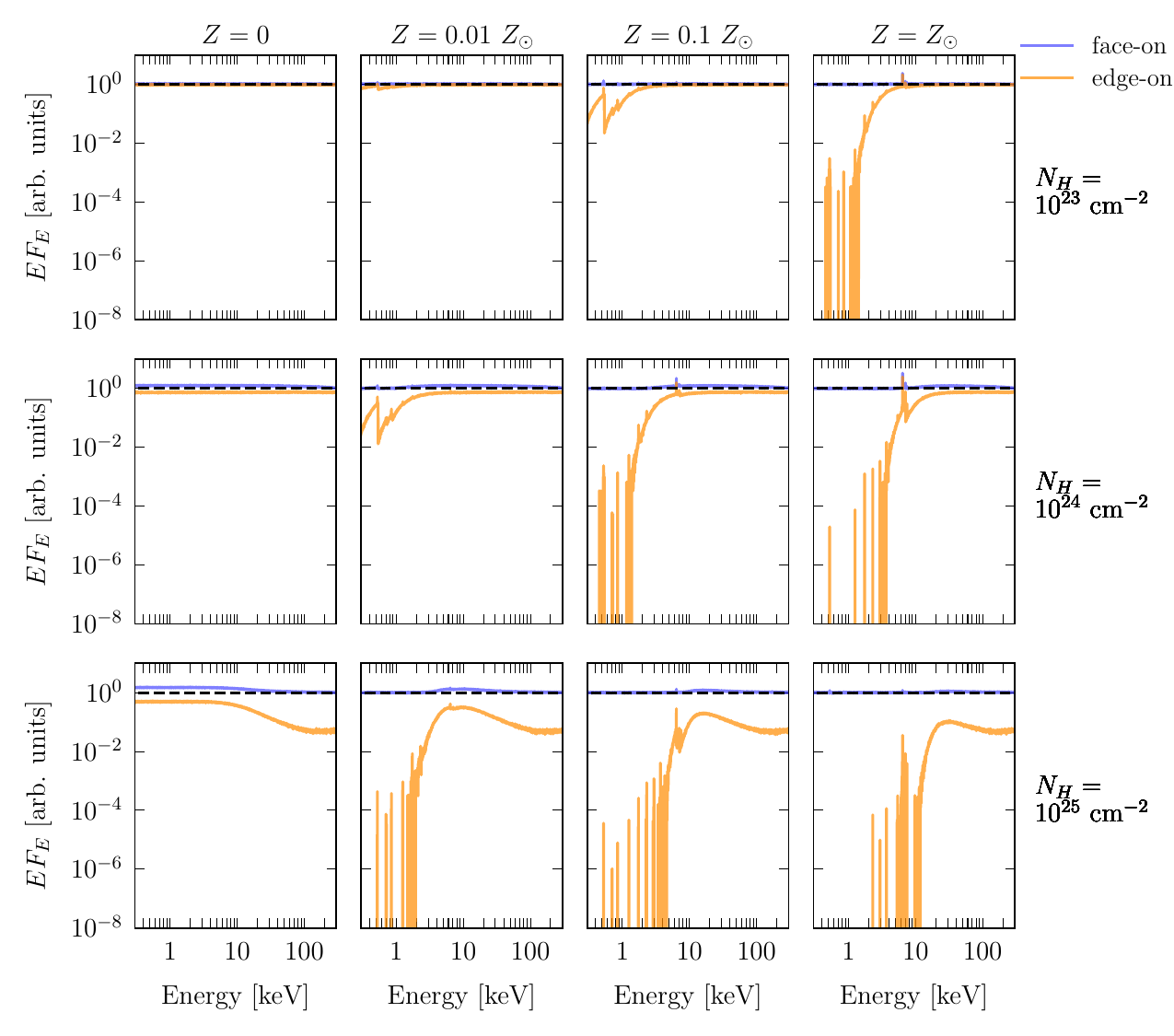}
    \caption{Comparison of edge-on and face-on spectra for increasing metallicity (left to right) and increasing column density (top to bottom). We assume $\Gamma = 2$ and $\theta_{\text{OA}} = 120^{\circ}$. All spectra are normalized to the incident power-law, shown with the dashed black line.}
    \label{fig:panel_spectra_120deg}
\end{figure*}

\begin{figure*}[h!]
    \centering
    \includegraphics[width=\textwidth,height=\textheight,keepaspectratio]{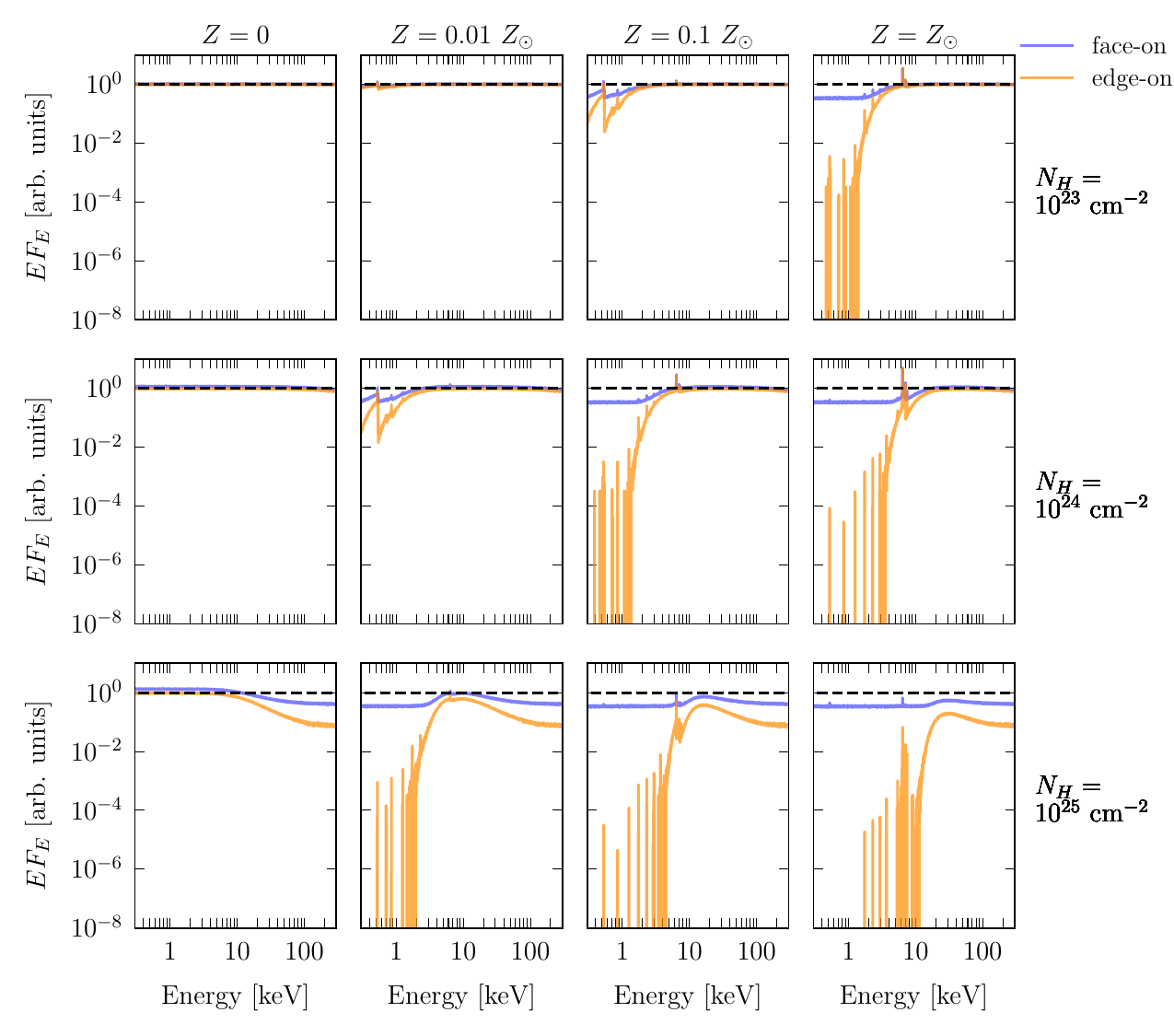}
    \caption{Same as Figure \ref{fig:panel_spectra_120deg} for an opening angle $\theta_{\text{OA}} = 30^{\circ}$.}
    \label{fig:panel_spectra_30deg}
\end{figure*}

The spectra we produce with our Monte Carlo runs cover an extensive range of parameters, shown in Table \ref{tab:param_space}. In addition to this, each run provides us with ten viewing angles and the ability to adjust the source power-law index, the redshift, and the intrinsic luminosity. In an attempt to distill this large amount of information, we will fix $\Gamma = 2$ and plot all spectra in the rest frame of the source (in Figures \ref{fig:basic_spectrum}, \ref{fig:panel_spectra_120deg}, \ref{fig:panel_spectra_30deg}, and \ref{fig:spectra_nonsolar}). We will also do away with physical units for the time being, focusing on the shape of the spectra in arbitrary $EF_E$ ($\nu F_{\nu}$) units, the escape fraction of photons, and equivalent width. When we discuss observations in Section \ref{subsec:spec_sims}, we will further explore a more extended parameter space. 

In Figure \ref{fig:basic_spectrum}, we show a typical edge-on spectrum of an obscured AGN with $N_H = 10^{24}$ cm$^{-2}$. One can make out a strong Fe K$\alpha$ doublet and fluorescent lines from several other species included in our model. The iron edge takes on its distinct shape upwards of $\sim7$ keV. In the right panel we can distinguish between the Fe K$\alpha_1$ and Fe K$\alpha_2$ lines at 6.404 and 6.391 keV, respectively. A prominent Compton shoulder appears to the red side of the Fe K$\alpha$ complex, as described in Section \ref{subsec:eqw}. Monte Carlo noise is evident in both panels, though we bin our spectra logarithmically in energy to mitigate this effect where photons are scarce. Now, we will turn to examining trends for these spectra as we vary certain parameters.

Figure \ref{fig:panel_spectra_120deg} shows face-on (Bin 1) and edge-on (Bin 10) spectra for the $\theta_{\text{OA}} = 120^{\circ}$ case for varying column density $N_H$ and metallicity. Moving from left to right, increasing the metallicity has the expected effects on edge-on sources: photoelectric absorption removes a significant portion of the soft X-rays and fluorescent lines begin to appear. The face-on spectra show subtler changes. For zero metallicity (pure Compton scattering) and $N_H \geq 10^{24}$ cm$^{-2}$, neighboring lines of sight scatter soft photons into the face-on bin and slightly lift the power-law above the input spectrum. As we drive metallicity upwards, those photons are absorbed within the torus and the flux below $\sim10$ keV comes mostly from direct emission into the bin. 

In the most heavily obscured sources (i.e. $N_H \gtrsim 10^{24}$ cm$^{-2}$, prominent features include the Compton hump and the Fe K$\alpha$ line. The Compton hump results from the repeated down-scattering of hard X-rays with $E \gtrsim 10$ keV, which on average lose more energy per scattering than softer photons. In the pure scattering panel on the bottom-left of Figure \ref{fig:panel_spectra_120deg}, we see the right side of the hump beginning to appear beyond 10 keV. Moving to higher values of $Z$, the full hump-like shape emerges. With nonzero metallicity, the Fe K edge defines its left boundary, as the spectrum then slopes upward due to the decreasing opacity of iron. This is consistent with observations, where the Compton hump peaks between 20 and 30 keV.

Figure \ref{fig:panel_spectra_30deg} shows the case of a $30^{\circ}$ torus opening angle. Here, even the ``face-on" bin is partially covered by the obscuring torus. Now, spectra show absorption below the input power-law in the soft X-rays, whereas in Figure \ref{fig:panel_spectra_120deg} the face-on viewing angle could only \textit{gain} flux from other bins. The Compton hump and Fe K$\alpha$ line rise higher than their counterparts in the $\theta_{\text{OA}} = 120^{\circ}$ case. We can understand this as the contribution of scattering from \textit{all} solid angle bins. In this geometry, very few photons leak out of the top and bottom openings; the covering fraction is greater than 96\%. Thus a greater number of hard photons will undergo Compton scattering and reduce their energies, resulting in a stronger hump.

To contextualize high-redshift X-ray observations, we will now examine the fraction of energy flux ($EF_E$) transmitted in some fixed band relative to what was emitted by the X-ray source into a particular line of sight. That is, $f_\text{trans} \equiv F_{\text{esc, band}} / F_{\text{emit, band}}$ for each solid angle bin. We do this as a function of redshift, keeping the observer's band fixed at 0.2-10 keV.

\begin{figure*}[h!]
    \centering
    \includegraphics[width=\textwidth,height=\textheight,keepaspectratio]{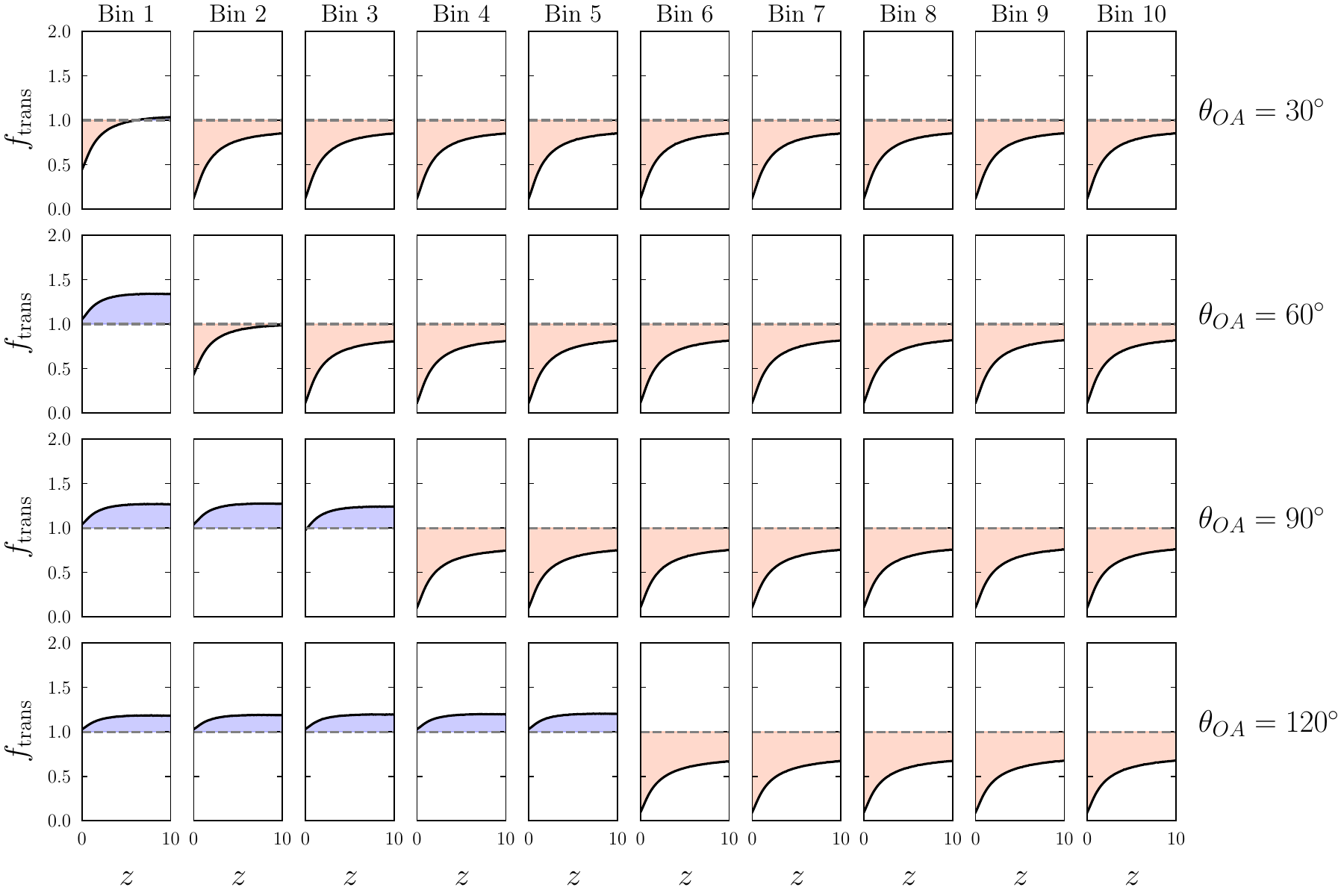}
    \caption{Fraction of energy flux ($EF_E$) transmitted in the 0.2-10 keV band as a function of redshift relative to the flux emitted into a solid angle bin (shown with the dotted gray line at $f_\text{trans} = 1$). Each row represents a different opening angle, increasing from top to bottom. Each column represents a different inclination angle, increasing from left to right (Bin 1 = face-on, Bin 10 = edge-on). Redshifting the source while keeping the observed frame energy band constant essentially increases the photon energies in the rest frame. This figure shows results for $Z = Z_{\odot}$ and $N_H = 10^{24}$ cm$^{-2}$.}
    \label{fig:obs_frac_Z_1}
\end{figure*}

\begin{figure*}[h!]
    \centering
    \includegraphics[width=\textwidth,height=\textheight,keepaspectratio]{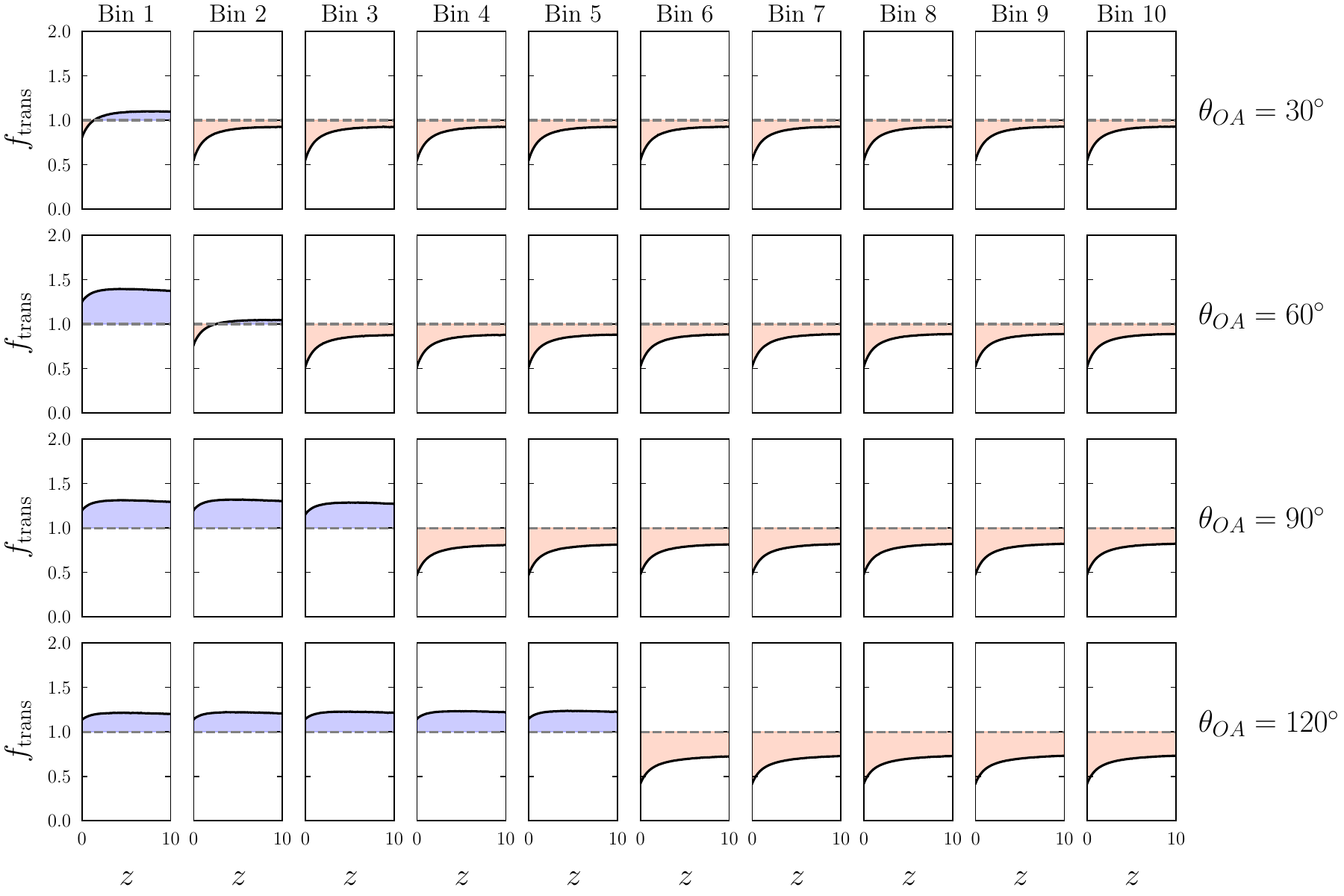}
    \caption{Same as Figure \ref{fig:obs_frac_Z_1} for $Z = 0.1 Z_{\odot}$, keeping column density fixed at $N_H = 10^{24}$ cm$^{-2}$.}
    \label{fig:obs_frac_Z_0.1}
\end{figure*}

\begin{figure*}[h!]
    \centering
    \includegraphics[width=\textwidth,height=\textheight,keepaspectratio]{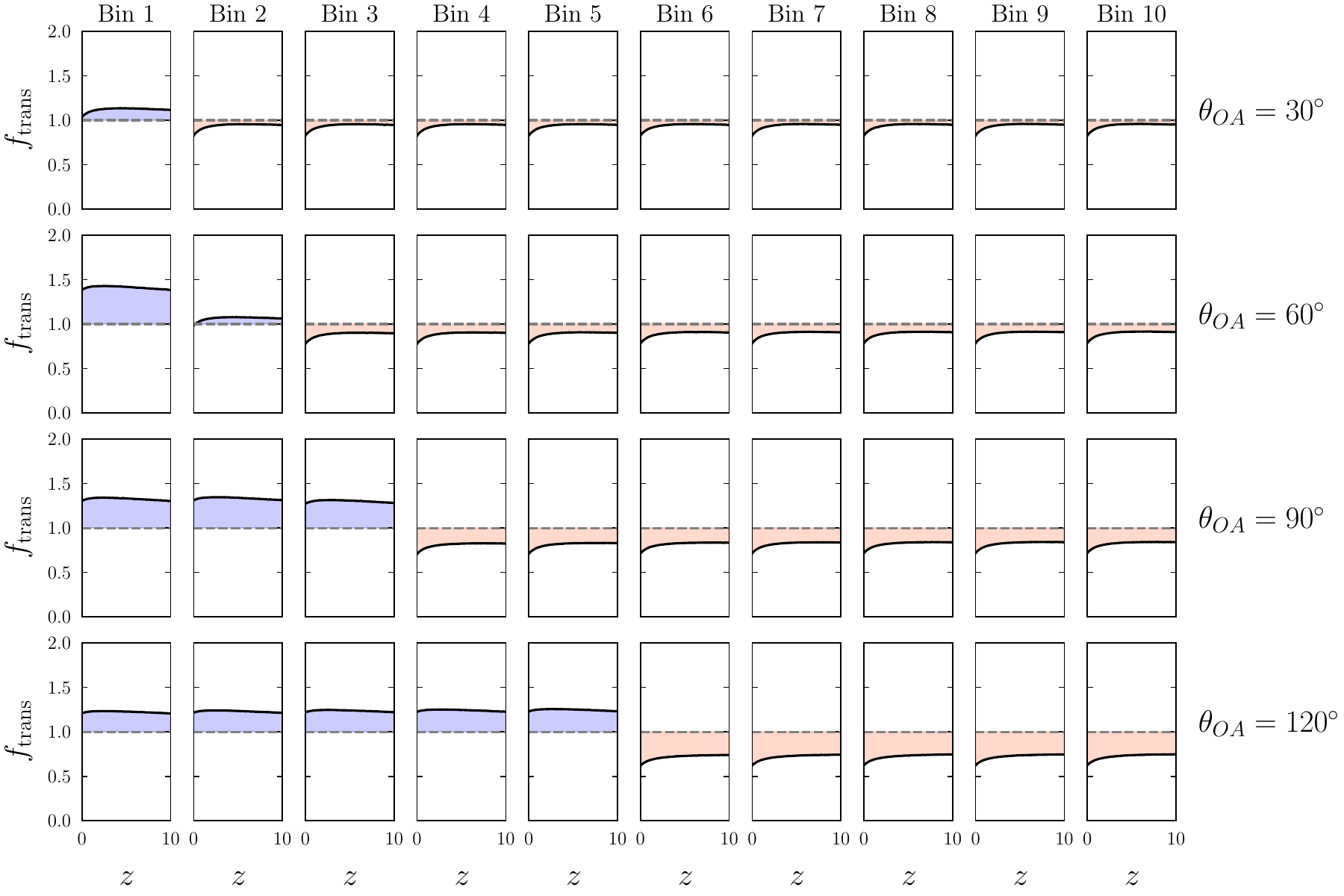}
    \caption{Same as Figure \ref{fig:obs_frac_Z_1} for $Z = 0.01 Z_{\odot}$, keeping column density fixed at $N_H = 10^{24}$ cm$^{-2}$.}
    \label{fig:obs_frac_Z_0.01}
\end{figure*}

\begin{figure*}[h!]
    \centering
    \includegraphics[width=0.95\linewidth]{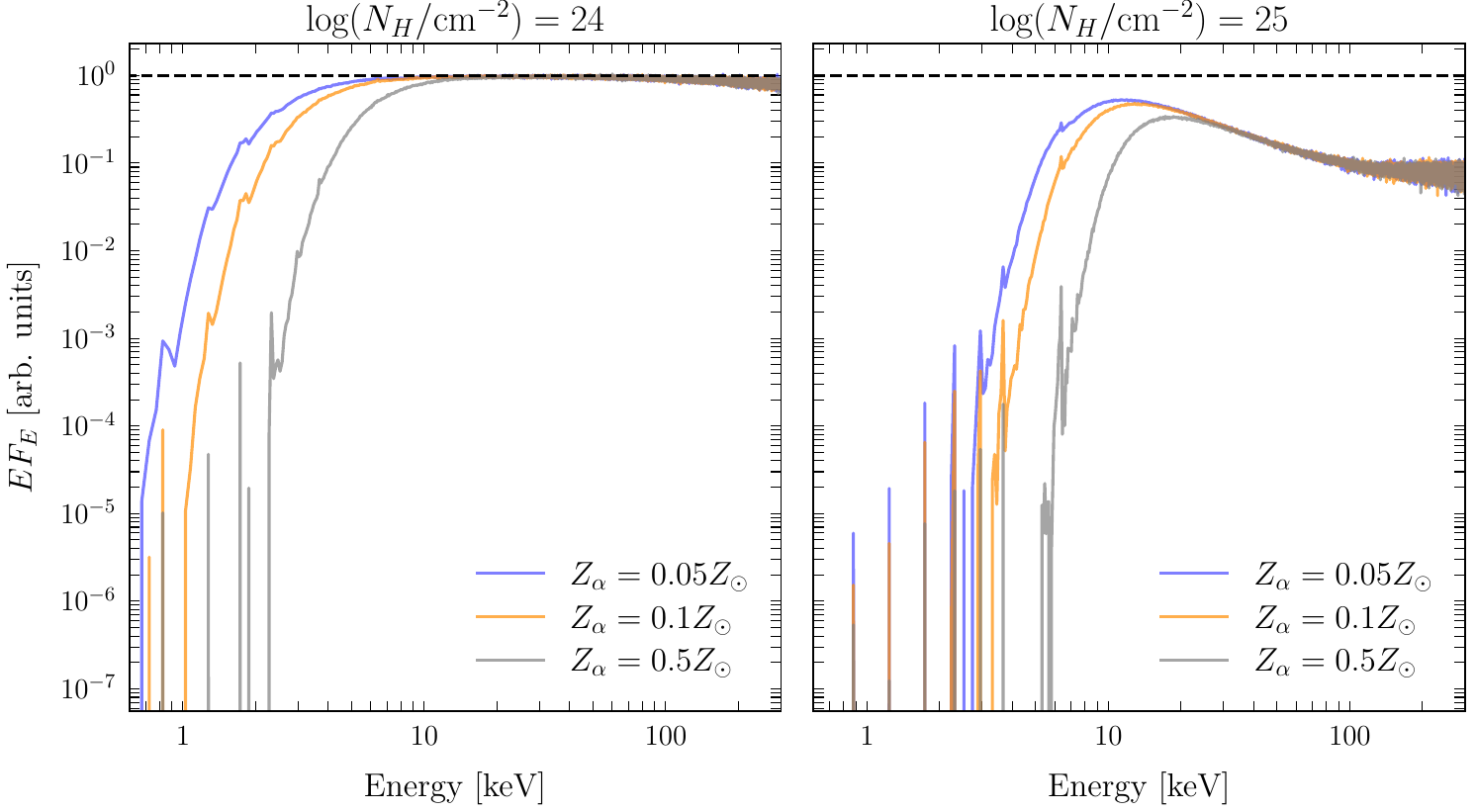}
    \caption{Comparison of spectra for non-solar abundance ratios. Like Figures \ref{fig:panel_spectra_120deg} and \ref{fig:panel_spectra_30deg}, the black dashed line indicates the input power-law. Each color represents a different abundance of the $\alpha$-elements, with iron-peak elements (Cr, Fe, Ni) held fixed at 1\% of their solar value. All spectra are for $\Gamma=2$, $\theta_{\text{OA}}=30^{\circ}$, and an edge-on viewing angle. We note that the shapes of the spectra are dependent on the opening angle in a similar sense to our comparison between Figures \ref{fig:panel_spectra_120deg} and \ref{fig:panel_spectra_30deg}.}
    \label{fig:spectra_nonsolar} 
\end{figure*}

Figures \ref{fig:obs_frac_Z_1}, \ref{fig:obs_frac_Z_0.1}, and \ref{fig:obs_frac_Z_0.01} show $f_\text{trans}$ as a function of redshift for $Z = Z_{\odot}$, $Z = 0.1 Z_{\odot}$, and $Z = 0.01 Z_{\odot}$, respectively, where $N_H = 10^{24}$ cm$^{-2}$ for all these figures. We emphasize here that each individual panel is less important than the trends across the entirety of the plot or between the three metallicities. For solar abundances in Figure \ref{fig:obs_frac_Z_1}, we can see that sightlines intercepting the torus immediately suffer from absorption. In contrast, unobstructed viewing angles \textit{gain} flux due to scattering from other bins, as we expect from our discussion of spectra in Figures \ref{fig:panel_spectra_120deg} and \ref{fig:panel_spectra_30deg}. Note the evolution of $f_\text{trans}$ with redshift. As we place the source at higher redshift, more photons emitted in the harder energy bands are able to escape, avoiding atomic absorption edges. Examining the rightmost (edge-on) column, we see another interesting effect: $f_\text{trans}$ is higher for smaller opening angles. This occurs for the same reason as the stronger Compton hump in Figure \ref{fig:panel_spectra_30deg}; fewer photons escape through the hollow regions, instead scattering into other bins. In the top-left panel, there is a ``turnover" point where the flux in the observed band transitions from being suppressed to enhanced at $z \sim 5$. This corresponds to softer photons getting absorbed while the harder energy bands still have contributions from other solid angle bins.  

Let us now tune the metallicity down to $0.1 Z_{\odot}$ in Figure \ref{fig:obs_frac_Z_0.1}. As we did before, we can focus our attention on the last column with an edge-on line of sight. Here, $f_\text{trans} \gtrsim 0.7$ at $z = 10$, with the most closed-up torus showing $f_\text{trans} \gtrsim 0.9$ for this viewing angle. Repeating this once again for a metallicity of $0.01 Z_{\odot}$, Figure \ref{fig:obs_frac_Z_0.01} shows that the transmitted flux fraction at high redshift approaches unity for most inclinations in the $\theta_\text{OA} = 30^{\circ}$ case and remains above about 0.75 for all covering fractions. These values of $f_\text{trans}$ may give us an optimistic outlook for X-ray observability. 

% However, high-redshift observations are challenging. We have not yet taken into account the detector nor various background sources. In Section \ref{subsec:observations}, we will perform realistic simulations which include these. 

We also test the effects of non-solar abundance ratios, as noted in Table \ref{tab:param_space}. This is motivated by the DTD of Type Ia supernovae, from which most of the iron-peak elements originate in the Universe. In these runs, we hold the iron-peak elements fixed at 1\% of their solar abundance and allow the remaining $\alpha$-element abundances (denoted by $Z_\alpha$ in the following figures) to vary together.

Figure \ref{fig:spectra_nonsolar} shows a sample of some of these spectra. As in Figures \ref{fig:panel_spectra_120deg} and \ref{fig:panel_spectra_30deg}, these are normalized to the input power-law at $EF_E=1$. We focus here on very high column densities, assuming $\Gamma=2$, $\theta_{\text{OA}}=30^{\circ}$, and looking at the system edge-on through the torus. From this figure, it is evident that the abundance of $\alpha$-elements drives photon absorption at soft energies. The attenuation is higher in this regime compared to the second

\clearpage

\noindent column of Figure \ref{fig:panel_spectra_30deg}, where all metals are fixed at $0.01 Z_{\odot}$. At $N_H = 10^{25}$ cm$^{-2}$, the continuum is sufficiently suppressed so that the Fe K$\alpha$ line and Fe K edge begin to appear. Hard X-rays above $\gtrsim20$ keV are unaffected by the variation in $Z_\alpha$. 

\begin{figure*}
    \centering
    \includegraphics[width=0.65\linewidth]{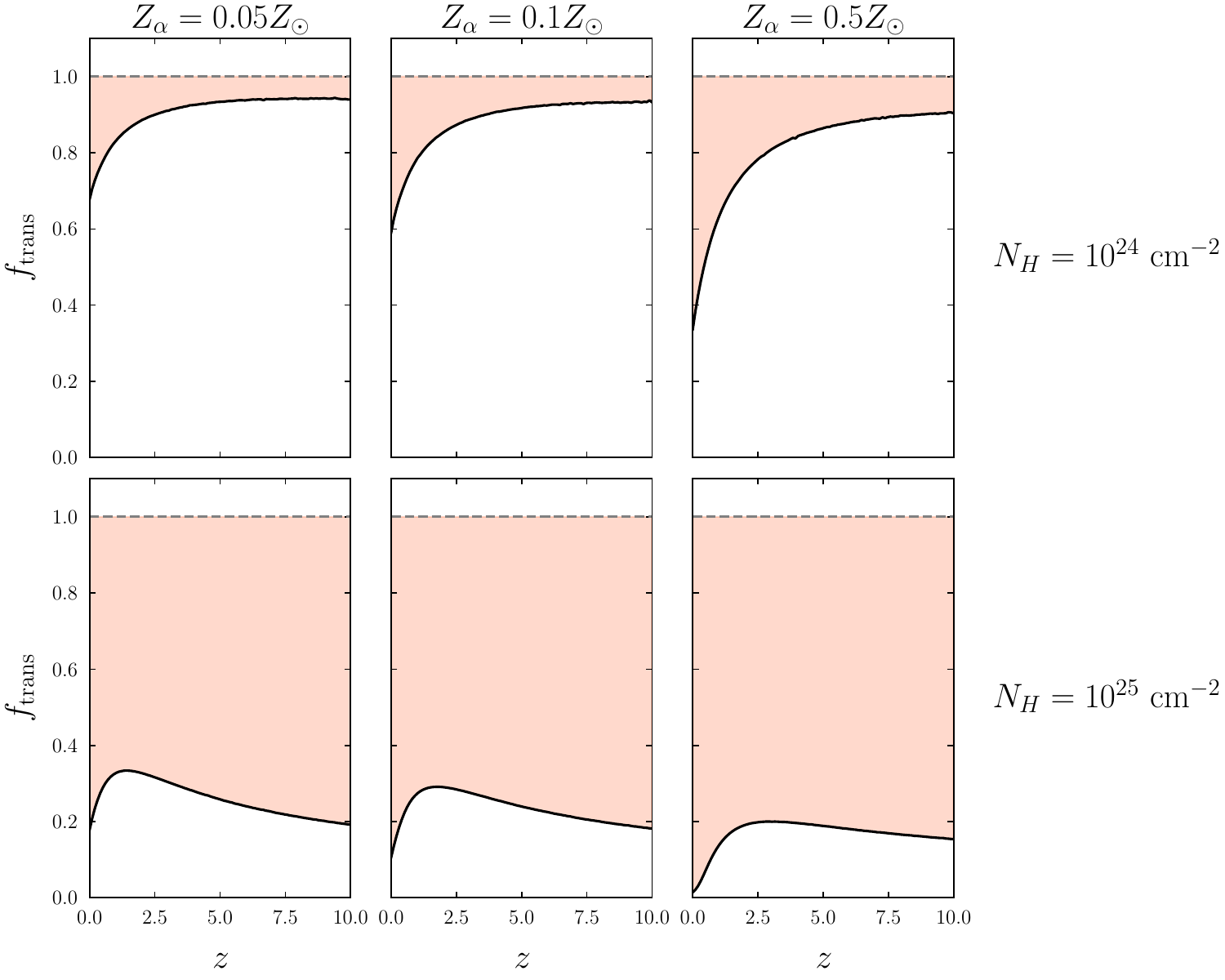}
    \caption{Transmitted flux fraction as a function of redshift for the cases shown in Figure \ref{fig:spectra_nonsolar} ($\Gamma=2$, $\theta_{\text{OA}}=30^{\circ}$, edge-on viewing angle). Iron-peak elements (Cr, Fe, Ni) are held fixed at 1\% of their solar value. $f_\text{trans}$ has the same definition as in Figures \ref{fig:obs_frac_Z_1}-\ref{fig:obs_frac_Z_0.01}.}
    \label{fig:obs_frac_nonsolar}   
\end{figure*}

To complement Figure \ref{fig:spectra_nonsolar}, we also plot $f_\text{trans}$ for all six spectra in Figure \ref{fig:obs_frac_nonsolar}. We see that the most pronounced changes in the transmitted flux occur at low redshift, when the soft X-ray band contributes the majority of photons. At high redshift, these values of $f_\text{trans}$ converge somewhat, though lower $Z_\alpha$ results in a modestly larger escape fraction. We stress that even for extreme column densities of $10^{25}$ cm$^{-2}$, $\gtrsim15\%$ of the flux emitted in our observable band escapes in the $z=10$ rest frame. Since departures from solar abundance ratios are more important at earlier times, we conclude that the effects of a Type Ia DTD will not be apparent when observing high-redshift AGNs.

\subsection{Strength of the Fe K$\alpha$ Line} \label{subsec:feka_strength}
We now turn to a discussion of the Fe K$\alpha$ line strength. We describe our method for calculating the equivalent width in Section \ref{subsec:eqw}. As with other metrics discussed thus far, the EW depends on the line of sight, column density, $\Gamma$, metallicity, and covering fraction of the obscuring torus. We will discuss all of these factors except for $\Gamma$, again fixing it at 2. 

\begin{figure*}
    \centering
    \includegraphics[width=1.0\linewidth]{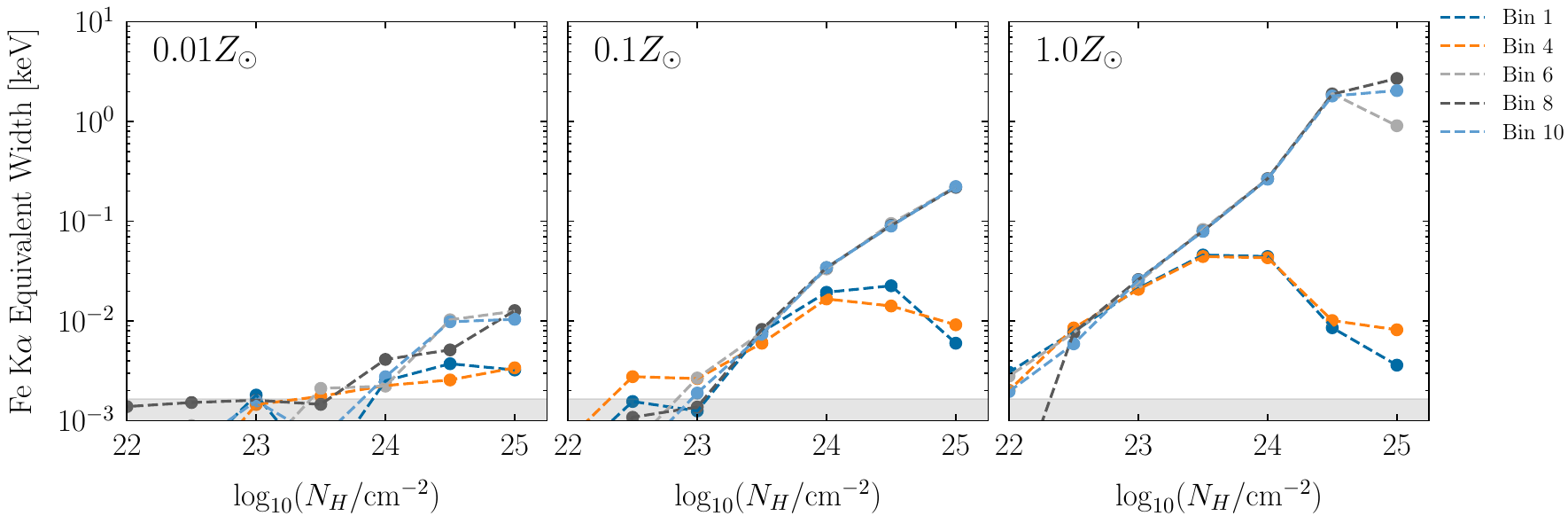}
    \caption{The equivalent width of the Fe K$\alpha$ line in selected solid angle bins as a function of line-of-sight column density for $\Gamma = 2$ and $\theta_{\text{OA}} = 120^{\circ}$. The leftmost panel shows $Z = 0.01 Z_{\odot}$, the middle panel shows $Z = 0.1 Z_{\odot}$, and the rightmost panel shows $Z = Z_{\odot}$. The gray shaded region represents the noise level calculated in Section \ref{subsec:eqw} and shown in Figure \ref{fig:eqw_noise}. Values in this region should be interpreted with care.}
    \label{fig:eqw_120deg}
\end{figure*}

\begin{figure*}
    \centering
    \includegraphics[width=1.0\linewidth]{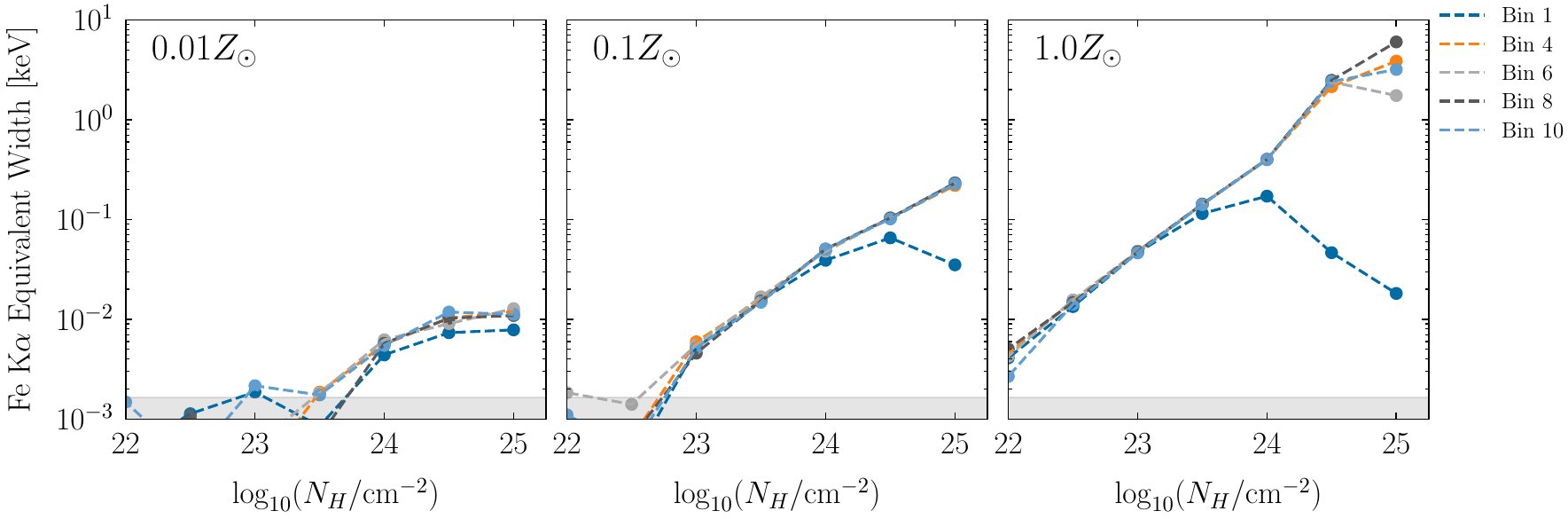}
    \caption{Same as Figure \ref{fig:eqw_120deg}, but for a torus with $\theta_{\text{OA}} = 30^{\circ}$. $\Gamma = 2$ as before.}
    \label{fig:eqw_30deg}
\end{figure*}

Figures \ref{fig:eqw_120deg} and \ref{fig:eqw_30deg} show the progression of the EW across metallicity and between two extremes of our geometry, $\theta_{\text{OA}} = 120^{\circ}$ and $\theta_{\text{OA}} = 30^{\circ}$, respectively. In both figures, the EW shows a clear and expected trend with metallicity. EWs are highest when photons have the largest interaction probability with atoms, in this case with iron. 

In Figure \ref{fig:eqw_120deg}, we see a general increase with $N_H$ in line with Figure 8 of MY09. Viewing angles that intercept the torus distinguish themselves from those that do not at $Z = Z_{\odot}$; EW $\gtrsim 1$ keV at the highest column density. For lower inclinations, the EW reaches a plateau or drops off around $N_H = 10^{24}$ cm$^{-2}$, as more photons are absorbed before they can produce a line and scatter into these bins. 

Figure \ref{fig:eqw_30deg} shows a similar trend. The EWs for intermediate to high inclinations are of order a few keV at solar metallicity, though more lines of sight can result in high EWs due to the large covering fraction. This means that observationally, there is a degeneracy between opening angle and inclination; measuring a large Fe K$\alpha$ EW does not necessarily go hand-in-hand with an edge-on viewing angle if we believe such extreme covering fractions are possible.

Lastly, we note that the calculated Fe K$\alpha$ EWs for the non-solar abundances are higher than the $Z = 0.01 Z_{\odot}$ case by up to factors of $\sim70$, with dependence on the covering fraction and column density. Since the iron abundance is held fixed and the continuum is identical for all our runs, the only remaining explanation is the prominent absorption from calcium. For iron line photons of energy 6.4 keV, the Ca cross section is 0.0187 Mb. For comparison, the closest iron-peak edge is from Cr, for which the cross section is 0.0336 Mb at the same energy. This is less than a factor of 2 higher than Ca \citep{verner_analytic_1995}. Additionally, the abundance of Ca is already a factor 4 higher than that of Cr when in solar ratios \citep{lodders_solar_2025}. When $\alpha$-element abundances are 5, 10, or 50 times higher than the iron-peak elements, Ca can dominate over Cr in absorbing fluorescent iron emission. Thus, the increase in Ca further diminishes the continuum at 6.4 keV, producing a measurably larger Fe K$\alpha$ EW for our non-solar ratios.

\section{Discussion} \label{sec:discussion}

\subsection{Observational Prospects for Low-Metallicity AGNs} \label{subsec:obs_prospects}
We have shown in the preceding section that metallicity indeed has a sizable effect on the X-ray reprocessing properties of AGNs. In Figures \ref{fig:panel_spectra_120deg} and \ref{fig:panel_spectra_30deg}, we can see that in CT, edge-on sources with $N_H \sim 10^{24}$ cm$^{-2}$, the energy flux is depressed by at most a factor of a few above 10 keV for solar metallicity. Reducing the abundances by an order of magnitude means that the source spectrum is relatively undiminished in the band above 2 keV. This is reflected in Figure \ref{fig:obs_frac_Z_0.1}, where at least half of the flux escapes at $z = 10$ (corresponding to rest-frame energies of 2.2 - 110 keV) in the rightmost column of panels. If one were to look at a lensed field, magnifications like that of UHZ1 ($\mu$ = 3.81; \citealt{bogdan_evidence_2024}) would eliminate this modest loss. 

One can find an interesting geometric dependence on this flux reduction: closed-up tori result in more photons escaping from the edge-on viewing angle as compared to tori with larger opening angles. This effect becomes more apparent as we move through Figures \ref{fig:obs_frac_Z_1}, \ref{fig:obs_frac_Z_0.1}, and \ref{fig:obs_frac_Z_0.01} in order, once again focusing on the rightmost column. At the lowest metallicities and $\theta_\text{OA} = 30^{\circ}$, there is barely a reduction in photon counts at high redshift. Though this may not be intuitive, there is significant geometric beaming from low to higher inclinations. In open geometries, more photons immediately escape through the funnel region without scattering and cannot replace those lost to absorption in other bins. 

% Because of this, there may be an unresolvable degeneracy between inclination and column density when analyzing spectra from observations. - not sure if we want to say this

Hope is not lost for observing AGNs with even higher obscuration, at $N_H \sim 10^{25}$ cm$^{-2}$. A strong Compton hump peaks between 10-30 keV, and depending on the metal abundance, lies at most an order of magnitude below the continuum in energy flux. Similarly to the lower column density discussed previously, the Compton hump's strength increases with a smaller toroidal opening angle. Additionally, the spectral plots in Section \ref{sec:results} show that the peak of the Compton hump shifts to the right with larger values of metallicity. As the metal abundance grows larger, the neutral Fe K edge suppresses the emission further, and the ``recovery" back to the continuum occurs at higher energies. In principle, it may be possible to put limits on the metallicity by observing where the peak is in the spectrum. In practice, even making a robust detection of such heavily enshrouded AGNs would require high signal-to-noise and therefore a very sensitive instrument.

Though this work relies on lower abundances in the early Universe, the rate of enrichment at these times is debated and highly uncertain. Metallicity measurements from \textit{JWST} at high redshift find that enrichment may happen quickly and more efficiently than previously thought (e.g. \citealt{nakane_fe_2025}, \citealt{hsiao_jwst_2024}). Additionally, the dense environment around an AGN may also experience chemical evolution at a higher rate than the rest of the host galaxy (\citealt{artymowicz_star_1993}, \citealt{hamann_chemical_1993}, \citealt{groves_emission-line_2006}). Other work suggests that there may be flat metallicity gradients in high-redshift galaxies, especially at early stages of bursty star formation and frequent mergers (e.g. \citealt{maiolino_re_2019}). Therefore, it is important to examine the effects of low metallicity when studying objects so early in the history of the Universe.

In particular, there is a lack of reliable iron abundance measurements at high redshift. The best \textit{JWST} detection of Fe II emission at intermediate redshifts was made at $z=3.5$ but still required deep observations assisted by gravitational lensing \citep{kokorev_deepest_2025}. There are some early indications of an empirical relation between iron abundance and quasar luminosity for $z>5$ \citep{trefoloni_missing_2025}. However, pushing beyond $z=6$ is crucial for quantifying the effects of a DTD for Type Ia supernovae. It is at these early epochs that we could expect to see $\alpha$-element enrichment without similar levels of iron. The only example of a significant iron detection in the first billion years comes from \citet{lambrides_discovery_2025}. They found several significant, multiply-ionized iron emission lines in a \textit{JWST} spectrum of a Little Red Dot (LRD; \citealt{matthee_little_2024}) at $z=6.68$. Similar observational work in the future may reveal how each elemental abundance evolves individually over cosmic time.

\subsection{Spectral Simulations for \textit{Chandra} and \textit{AXIS}} \label{subsec:spec_sims}
With methods outlined in Section \ref{subsec:sim_obs}, we generate mock spectra for \textit{Chandra}/ACIS-I and \textit{AXIS}, comparing the observational prospects for each of these telescopes at $z=8$ and $z=10$. We consider two source classes to simulate. First, we examine the case of an AGN with an intrinsic 2-10 keV luminosity of $10^{43}$ erg s$^{-1}$. The heavy black hole seeding scenario (e.g. \citealt{inayoshi_assembly_2020}) would produce several such sources detectable by deep \textit{AXIS} surveys. Our second source class has $L_{2-10} = 10^{44}$ erg s$^{-1}$. This X-ray luminosity is unexpected at such early times, as the black holes powering the AGNs would have to be unreasonably massive. However, \textit{JWST} results are already challenging the standard black hole seeding and growth paradigm. We include this high-luminosity class to clearly show the consequence of folding metallicity effects through a detector response.

We use a \textit{Chandra} exposure time of 7 Ms to match the deepest exposure it has taken, the CDFS. We include neither an instrumental nor an astrophysical background for \textit{Chandra}; this therefore represents the most favorable outlook for its performance. In contrast, we use both background sources when simulating \textit{AXIS} spectra. We choose an integration time of 5 Ms, which was the planned \textit{AXIS} deep field exposure during its Phase A study.

\begin{figure*}[ht!]
    \centering
    \includegraphics[width=\linewidth]{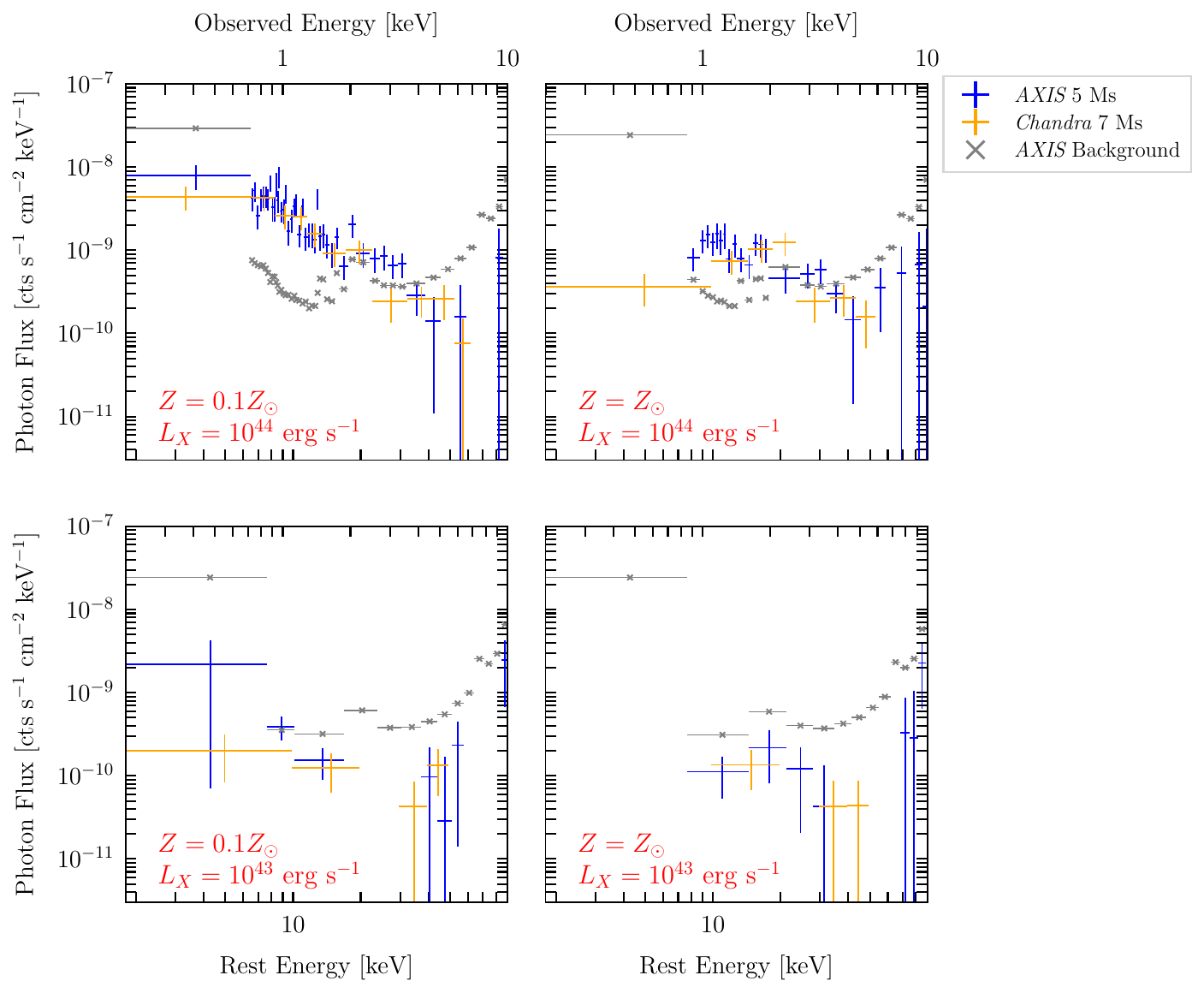}
    \caption{Simulated \textit{AXIS} and \textit{Chandra} spectra at $z=8$ generated with the \texttt{fakeit} command within \textsc{xspec}. Energies on the lower horizontal axis are in the rest frame of the source, while upper axes show the observed energy. The spectra have been re-binned for visualization purposes, with a minimum significance of 3$\sigma$ and maximum of 75 channels binned together. All sources shown here are viewed edge-on and have the following parameters: log($N_H$/cm$^{-2}$) = 24, $\theta_{\text{OA}} = 120^{\circ}$, $\Gamma = 2$. Metallicities and 2-10 keV X-ray luminosities are shown in red on each panel.}
    \label{fig:fakeit_z8}
\end{figure*}

Figure \ref{fig:fakeit_z8} shows \textit{AXIS} and \textit{Chandra} spectra of both luminosity classes at $z=8$, viewed edge-on for an obscuring column density of $10^{24}$ cm$^{-2}$. We also compare observed spectra for $Z = Z_{\odot}$ and $Z = 0.1 Z_{\odot}$ in the left and right columns, respectively. The bright, $L_X = 10^{44}$ erg s$^{-1}$ AGNs are easily detected by \textit{AXIS}. \textit{Chandra} would likely have made a detection of these sources, but as shown in Table \ref{tab:fakeit_counts}, ACIS-I would record an order of magnitude fewer photon counts than \textit{AXIS}. Metallicity effects are particularly evident when examining the spectra from these bright AGNs. Both telescopes receive significantly higher flux below $\sim10$ keV when abundances are reduced to $0.1 Z_{\odot}$.

The dimmer $L_X = 10^{43}$ erg s$^{-1}$ sources are background-dominated at all energies. However, with an accurate model of the background, a telescope like \textit{AXIS} could make a robust detection of these AGNs down to observed fluxes of $F_\text{0.5-2 keV} \sim \text{few} \times 10^{-18}$ erg s$^{-1}$ cm$^{-2}$ \citep{reynolds_overview_2023}. Even at solar metallicity (bottom right panel of Figure \ref{fig:fakeit_z8}), \textit{AXIS} would record several tens of X-ray photons. \textit{Chandra}, however, would be reaching the limit of its capabilities regardless of the metallicity. 

\begin{figure*}[ht!]
    \centering
    \includegraphics[width=\linewidth]{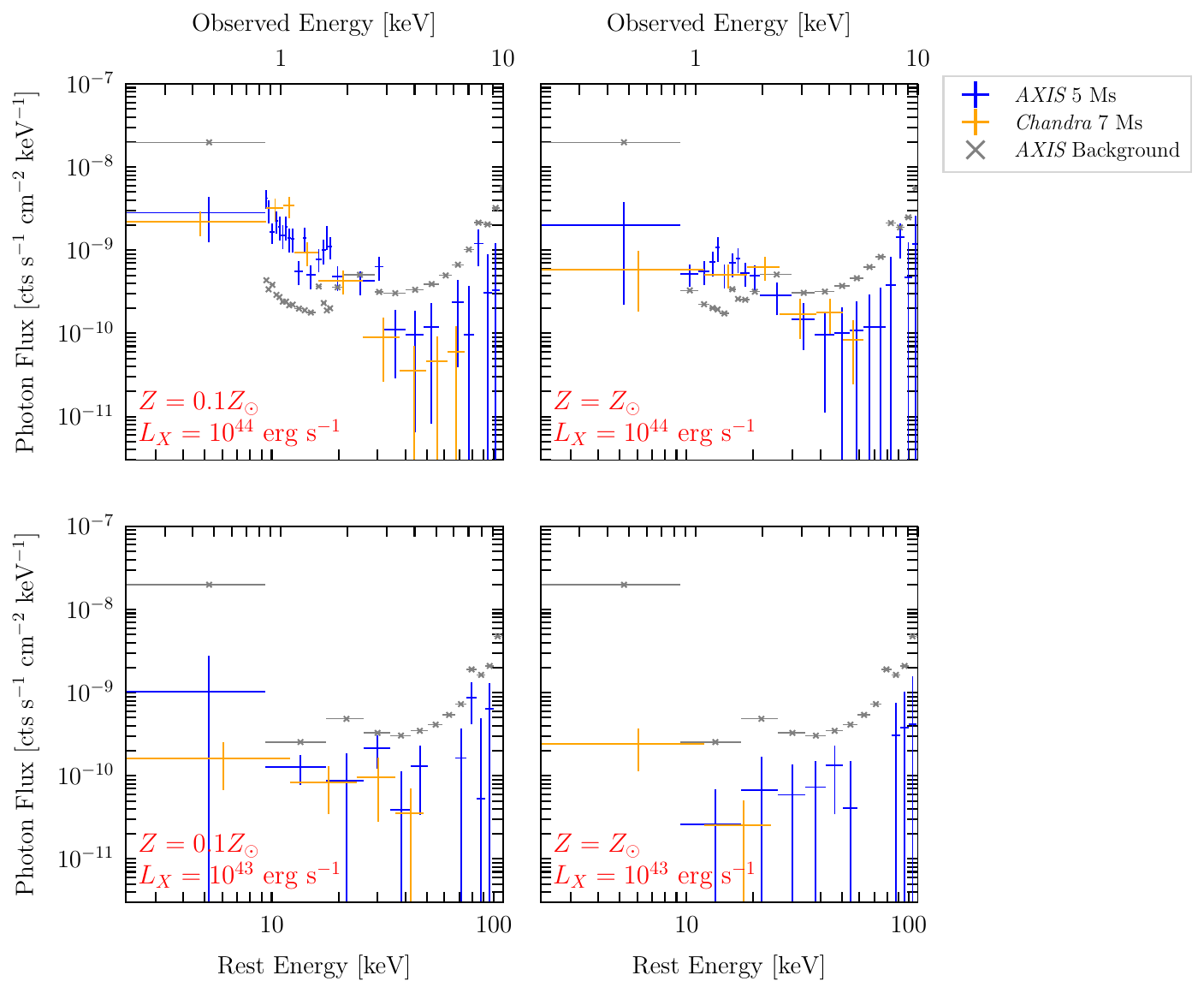}
    \caption{Same as Figure \ref{fig:fakeit_z8}, but all sources shown here have $z=10$. As before, $\theta_{\text{OA}} = 120^{\circ}$ and $\Gamma = 2$.}
    \label{fig:fakeit_z10}
\end{figure*}

In Figure \ref{fig:fakeit_z10}, we present the same luminosities and metallicities but at $z=10$. Again, the brighter sources are detected by both observatories. It is clear, however, that spectra from both instruments are sparser. Remarkably, \textit{AXIS} can still detect the AGNs with $L_X = 10^{43}$ erg s$^{-1}$. At solar metallicity, \textit{AXIS} will only receive $20-30$ photons, but could still make a detection due to the survival of hard X-rays above 10 keV in the rest frame. In contrast, luminosity is the determining factor for whether or not \textit{Chandra} can find these heavily enshrouded AGNs. ACIS-I would register at most a few photons from the dimmer sources.

From Table \ref{tab:fakeit_counts}, we find that the \textit{Chandra} photon counts are largely independent of metallicity, instead reflecting the luminosity of the source. This reinforces the 

\noindent fact that \textit{Chandra} is mainly sensitive to the hard X-ray emission, missing most of the photons below 10 keV. The degree to which \textit{AXIS} can find and characterize these AGNs is a stronger function of metal abundance. Since an \textit{AXIS}-like observatory would recover and improve upon \textit{Chandra}'s effective area below 1 keV, soft X-rays would constitute a sizable fraction of its observable photons. Indeed, for $z=10$, \textit{AXIS} may not make a significant detection of a $L_X = 10^{43}$ erg s$^{-1}$ AGN without sub-solar metallicity.  

\begin{table}[]
    \centering
    \begin{tabular}{c|c|c}
        \hline 
        \multicolumn{3}{c}{$z=8$} \\
        \hline
        Metallicity & $10^{43}$ erg s$^{-1}$ & $10^{44}$ erg s$^{-1}$ \\
        \hline
        $Z = 0.1 Z_{\odot}$ & 74 (11) & 974 (87) \\
        $Z = Z_{\odot}$ & 43 (6) & 362 (51) \\
        \hline
        \multicolumn{3}{c}{$z=10$} \\
        \hline
        Metallicity & $10^{43}$ erg s$^{-1}$ & $10^{44}$ erg s$^{-1}$ \\
        \hline
        $Z = 0.1 Z_{\odot}$ & 77 (9) & 808 (59) \\
        $Z = Z_{\odot}$ & 26 (5) & 274 (35) \\
    \end{tabular}
    \caption{Total background-subtracted, source counts for \textit{AXIS} (\textit{Chandra}) in the 0.2-10 keV observed-frame band for each case shown in Figures \ref{fig:fakeit_z8} and \ref{fig:fakeit_z10}. The top portion of the table is for sources at $z=8$ and the bottom is for sources at $z=10$. The luminosities shown in column headers are from the coronal power-law in the rest-frame 2-10 keV band.}
    \label{tab:fakeit_counts}
\end{table}

Future X-ray observatories will enjoy great synergy with \textit{JWST} once launched, and can perhaps further probe the purported X-ray weakness of LRDs with their large effective areas. There have been several attempts to detect X-rays from LRDs in deep, archival \textit{Chandra} exposures, most of which only resulted in upper limits on the X-ray luminosity (e.g. \citealt{matthee_little_2024}, \citealt{kocevski_rise_2025}, \citealt{yue_stacking_2024}, \citealt{maiolino_jwst_2025}, \citealt{sacchi_chandra_2025}, \citealt{comastri_jwst-discovered_2026}). One explanation put forth by, for example, \citet{kocevski_rise_2025}, \citet{maiolino_jwst_2025}, and \citet{comastri_jwst-discovered_2026} is that heavy obscuration is the reason for the lack of X-rays. An interesting possibility is that the buildup of dense gas is itself a consequence of lower metallicity, and therefore a reduced coupling between outward radiation pressure and dust \citep{ishibashi_another_2025}. The spectral simulations shown in our work indicate that metal-poor, CT gas surrounding an AGN at $z=8$ should still result in a statistically significant detection by a telescope similar to \textit{AXIS}. If there is still no such detection made by future missions, one could argue that the assumed bolometric correction ($L_\text{bol}/L_X$) is even greater than previously thought; that is, LRDs are intrinsically X-ray weak, and extremely so. However, we note that the majority of LRDs have been found at $z<8$, when some non-negligible enrichment has already occurred. The work we present here is focused on an even earlier phase of SMBH evolution.

\section{Summary and Conclusions} \label{sec:conclusion}

In this work, we have investigated the effects of metallicity on the distant, cold X-ray reflection spectrum from AGNs at high redshift. The main conclusions are summarized here:

\begin{enumerate}
    \item Reducing the abundances of metals in the dusty torus surrounding the central engine preserves some soft X-ray flux below the iron K band, allowing more photons to leak through. These effects are quite dramatic when metallicity is reduced by two or more orders of magnitude.
    \item The covering fraction of the torus plays a role in the escape fraction of photons. For appreciable column density, a closed-up torus will retain more photons and redirect them toward edge-on lines of sight, boosting the observable emission as compared to open geometries where photons can leave through the large openings. 
    \item Non-solar abundance ratios--namely enhanced $\alpha$-elements relative to iron--do not significantly alter the hard X-ray properties of AGNs at high redshift. $\alpha$ elements dominate absorption at soft energies. For this reason, the $6-7$ keV continuum and therefore the equivalent widths of the Fe K$\alpha$ line are quite sensitive to $Z_\alpha$.
    \item Typical AGNs with $L_X = 10^{43}$ erg s$^{-1}$ surrounded by CT material with sub-solar metallicity can be reliably detected up to $z = 10$ by future missions like \textit{AXIS}. Elemental abundances play an important role at soft energies, and would determine whether or not such a telescope can observe these sources when $N_H > 10^{24}$ cm$^{-2}$. If even more luminous AGNs exist at these high redshifts, the potential of an \textit{AXIS}-like observatory to extract detailed spectra will be transformative.
\end{enumerate} 

As \textit{JWST} continues its march into the uncharted territory of the high-redshift Universe, the vital role of deep, arcsecond X-ray imaging has become clear. Multi-wavelength confirmation of black hole activity at $z\gtrsim10$ will allow us to better understand their evolution and connect these SMBHs to local AGNs. We have shown here that low elemental abundances are a key ingredient in detecting such systems. Early black holes growing in metal-poor environments should be detected by future instruments, building upon the legacy of \textit{Chandra} by advancing the high-redshift X-ray frontier even further.

\vspace{5pt}

\section*{Acknowledgments}
We thank the referee for their insightful comments, which greatly improved the manuscript. This work made use of the \textsc{astra} computing server, which is owned and maintained by the Department of Astronomy at the University of Maryland, College Park.

\begin{contribution}
Y.A.G. was responsible for writing and running the Monte Carlo code, as well as analyzing the output and drafting the manuscript. C.S.R. assisted with writing the code, guided the physical interpretation of the results, and processed \textit{Chandra} data in order to produce the response files used in Section \ref{subsec:spec_sims}. C.S.R. also contributed to and edited the manuscript.
\end{contribution}

\section*{Software}
\textsc{boost} C++ Library, \textsc{numpy} \citep{harris_array_2020}, \textsc{matplotlib} \citep{hunter_matplotlib_2007}, \textsc{scipy} \citep{virtanen_scipy_2020}, \textsc{xspec} \citep{arnaud_xspec_1996}, \textsc{astropy} \citep{astropy_collaboration_astropy_2022}




\bibliography{paper}{}
\bibliographystyle{aasjournal}



\appendix \label{sec:appendix}
\counterwithin*{equation}{section} % reset equation counter for A1, A2, ... B1, B2 ...
\renewcommand\theequation{\thesection\arabic{equation}}

\section{Comparison to Published Models} \label{subsec:app_compare}

\begin{figure*}
    \centering
    \includegraphics[width=\linewidth]{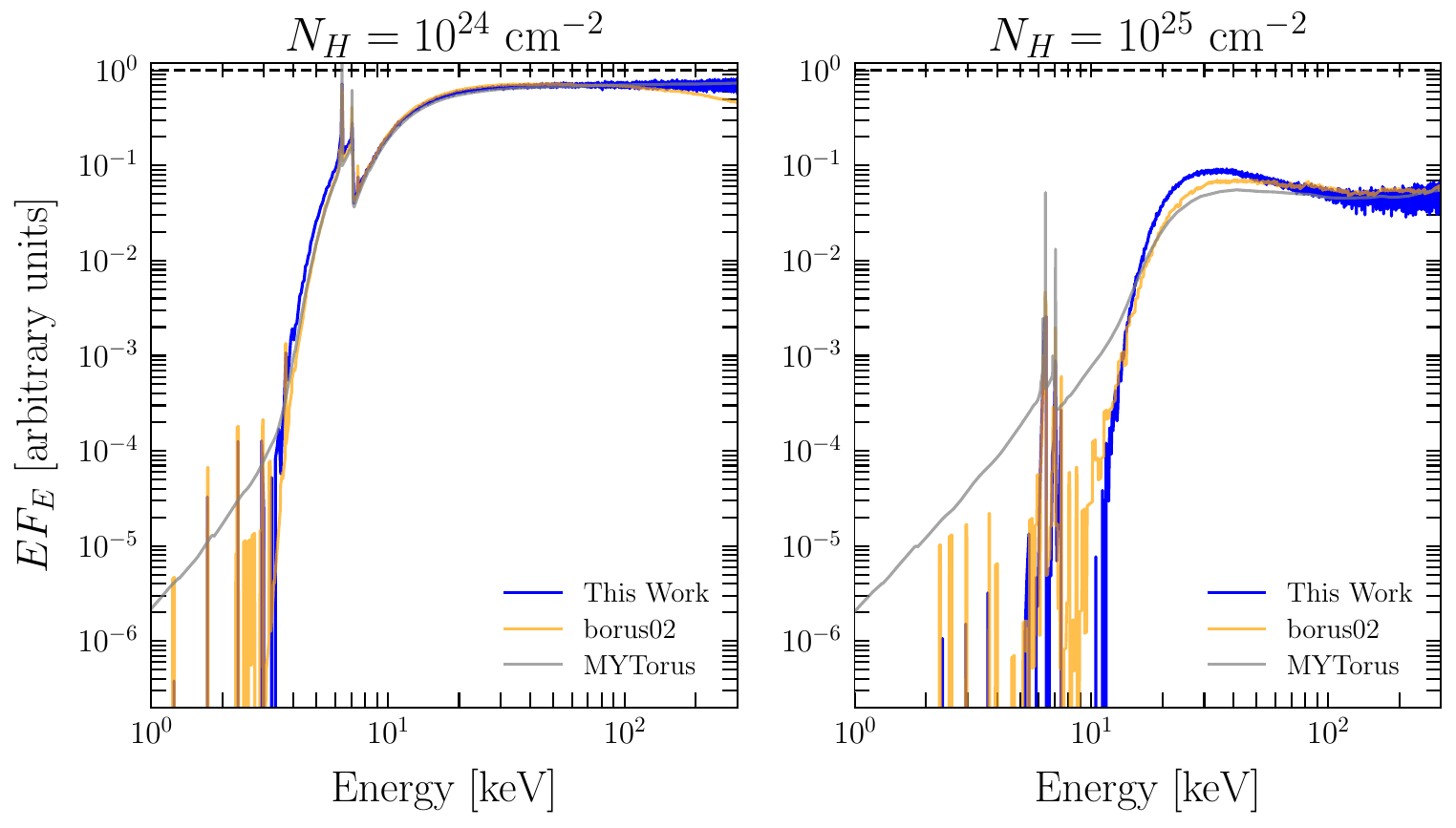}
    \caption{A comparison of our model (blue) to \textsc{mytorus} (gray) and \textsc{borus02} (orange) for solar metallicity, input power-law with $\Gamma = 2$, 50\% torus covering fraction, and edge-on viewing angle. The left panel shows models with $N_H = 10^{24}$ cm$^{-2}$ and the right panel shows those with $N_H = 10^{25}$ cm$^{-2}$.}
    \label{fig:modelcomparison_appendix}
\end{figure*}

To ensure our model captures all the relevant physics, we compared our spectra to the \textsc{mytorus} model \citep{murphy_x-ray_2009} and the \textsc{borus02} model \citep{balokovic_new_2018}. Our model has a spherical-toroidal geometry as illustrated in Figure \ref{fig:geometry} and detailed in Appendix \ref{subsec:app_geo}. \textsc{mytorus} uses a doughnut-like torus with circular cross sections that covers half the sky from the point of view of the X-ray source. The \textsc{borus02} geometry is a sphere with conical cutouts, which is quite similar to our work. In Figure \ref{fig:modelcomparison_appendix}, we show these comparisons for two column densities assuming edge-on viewing through the torus and a total opening angle of 120$^\circ$. The models qualitatively agree, with some key differences. 

The shape of the \textsc{mytorus} spectrum diverges from our model and \textsc{borus02} at energies below $\sim4$ keV for $N_H = 10^{24}$ cm$^{-2}$ and below $\sim15$ keV in the $N_H = 10^{25}$ cm$^{-2}$ case. This is because MY09 adopt a different approach to the low-energy regime. They argue that since absorption dominates here, the escape fraction at a given energy can be obtained by considering only the single-scattering albedo at that energy. They use their code to find escape fractions for a few albedo values and then interpolate between them when calculating their full Green's functions. Our code and \textsc{borus02} directly track photons in this regime without interpolation. 

For very high column densities, the three models have varying strengths and peaks of the Compton hump. This may be an effect of slight differences between them in the toroidal geometry. We also find that artificially considering photons to be absorbed once they scatter more than 15 times allows us to qualitatively match the \textsc{borus02} spectrum. However, the \textsc{borus02} model contains a known error in the convolution of the Green's functions with an input spectrum \citep{meulen_x-ray_2023}. This may partially explain the discrepancy in hard X-ray flux. In any case, all three models converge above 100 keV, since photons with these energies are least likely to interact with the medium. 

\textbf{As a final consideration}, we note that both \textsc{mytorus} and \textsc{borus02} use solar photospheric abundances from \citet{anders_abundances_1989}. For consistency, our spectra in Figure \ref{fig:modelcomparison_appendix} were calculated with these abundances as well. However our results in the main body of this paper were obtained using the abundances from the more recent work of \citet{lodders_solar_2025}. 

\section{Spherical-Toroidal Geometry} \label{subsec:app_geo}

\begin{figure}[hbt!] 
    \centering
    \includegraphics[width=0.75\linewidth]{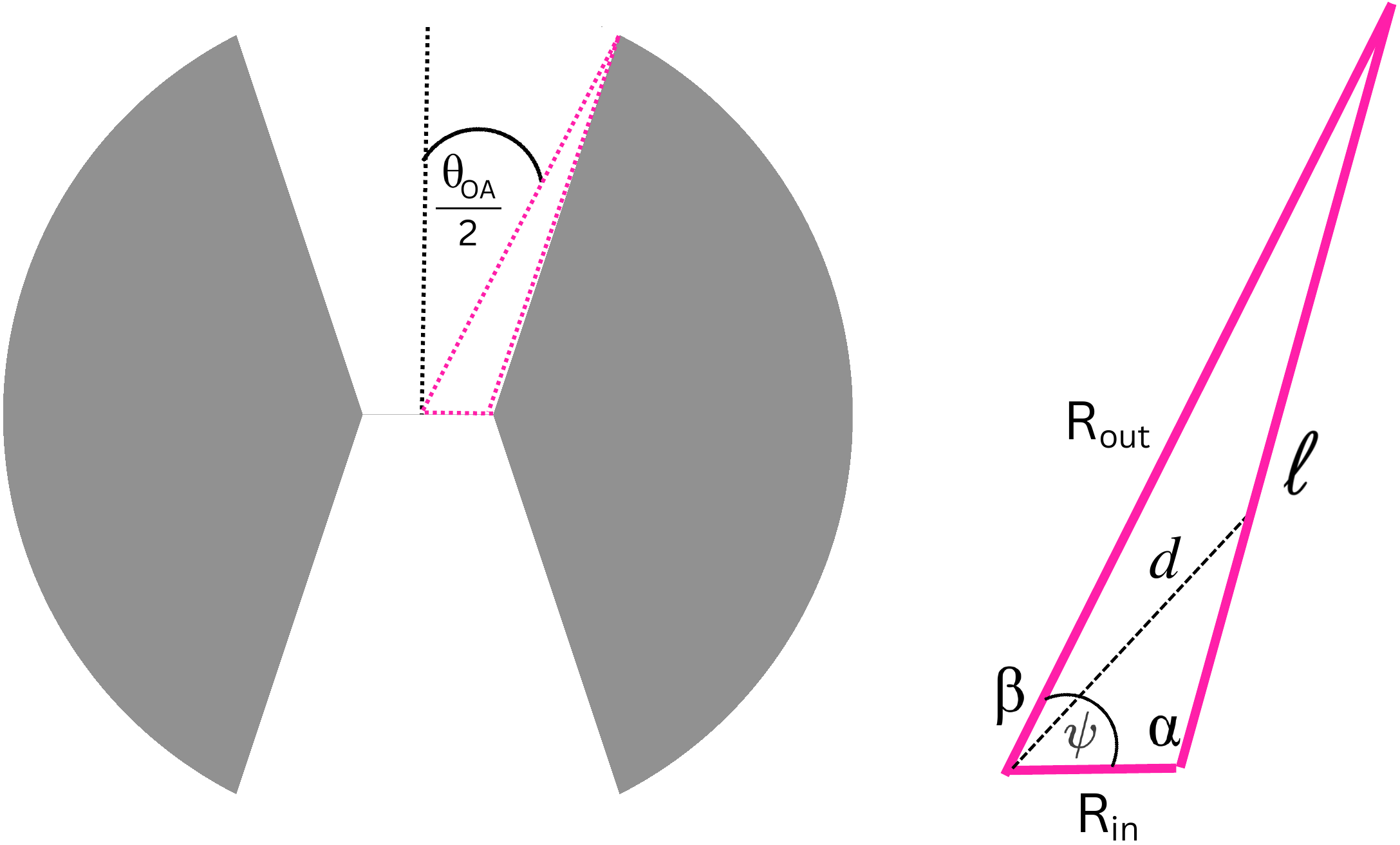}
    \caption{A view of the relevant geometry for determining whether or not photons are in the medium. The pink triangle on the right is a zoom-in of the dashed pink triangle on the left. $\theta_{\text{OA}}/2$ is the \textit{half}-opening angle of the torus.}
    \label{fig:geometry_appendix}
\end{figure}

Here, we will provide more details about the geometry used in our simulations. Let $\textbf{r} = (r, \theta, \phi)$ be the current position of a photon in spherical coordinates. Let $\psi \equiv \pi/2 - \theta$ and $\beta \equiv (\pi - \theta_{\text{OA}})/2 $ as in Figure \ref{fig:geometry_appendix}. Additionally, we can use the Law of Cosines to find that $\ell = (R_{\text{out}}^2 + R_{\text{in}}^2 - 2R_{\text{out}}R_{\text{in}}\text{cos} \, \beta)^{1/2}$. The problem we would like to solve is as follows. We want to find out when a photon is in the medium or outside of it. In order to do so, we must calculate $d$ as a function of $\psi$, which is the distance to the inner wall of the torus from the center given the photon's current value of $\psi$. If $d \leq r \leq R_{\text{out}}$, the photon is inside the medium. If $r < d$, the photon is in the hollow inner region. We will now derive the formula for $d(\psi)$. 

First, we notice that the pink triangle in Figure \ref{fig:geometry_appendix} itself can be split into two triangles by the line segment of length $d$. We call the smaller of these triangle A and the full pink triangle, triangle B. We use the Law of Sines on triangles A and B, respectively: 

\begin{subequations}
\begin{equation} \label{eq:law_sines_a}
    \frac{\text{sin} \, \alpha}{R_{\text{out}}} = \frac{\text{sin} \, \beta}{\ell}
\end{equation}
\begin{equation} \label{eq:law_sines_b}
    \frac{\text{sin} \, \alpha}{d} = \frac{\text{sin} (\pi - \psi - \alpha)}{R_{\text{in}}}
\end{equation}
\end{subequations}

\noindent By solving for and equating the sin$\alpha$ terms, we can see that 

\begin{equation} \label{eq:d_incomplete}
    d = \frac{R_{\text{in}} R_{\text{out}}}{\ell} \frac{\text{sin} \, \beta}{\text{sin} (\pi - \psi - \alpha)}
\end{equation}

\noindent Expanding sin$(\pi - \psi - \alpha)$ = sin$(\psi + \alpha)$ and recognizing that sin$ \, \alpha$ = $R_{\text{out}} \text{sin} \, \beta$/$\ell$ from Equation \ref{eq:law_sines_a}, we find 

\begin{equation} 
    \text{sin}(\psi + \alpha) = \text{sin} \, \psi \, \text{cos} \, \alpha + \text{cos} \, \psi \, \text{sin} \, \alpha = \text{sin} \, \psi \sqrt{1 - \left(\frac{R_{\text{out}} \text{sin} \, \beta}{\ell}\right)^2} + \text{cos} \, \psi\left(\frac{R_{\text{out}} \text{sin}  \, \beta}{\ell}\right)
\end{equation}

\noindent Note here that we have made use of the identity sin$^2\alpha$ + cos$^2\alpha$
 = 1 to ensure we eliminate all instances of $\alpha$ from the final result. At last, we arrive at our desired form for $d$: 

 \begin{equation} \label{eq:d_complete}
     d(\psi) = \frac{R_{\text{out}}R_{\text{in}}}{\ell} \text{sin} \, \beta \left[ \text{sin}  \, \psi \sqrt{1 - \left(\frac{R_{\text{out}} \text{sin}  \, \beta}{\ell}\right)^2} + \text{cos} \, \psi\left(\frac{R_{\text{out}} \text{sin}  \, \beta}{\ell}\right) \right]^{-1}
 \end{equation}

 We have now devised a strategy for checking whether a photon is inside the medium or not; we simply calculate $\psi$, use Equation \ref{eq:d_complete} to calculate $d$, and then compare to the photon's distance from the center of the torus, $r$. 

 We now lay out the escape conditions for each photon. A photon is deemed to have escaped the medium when one of the following is satisfied:
\begin{enumerate}
    \item The photon is initialized in a direction that will never intersect the medium. That is, the trajectory lies entirely within the ``cut-out" region defined by $\theta_{\text{OA}}$. 
    \item The photon exits the medium at an inner wall of the cut-out region and will not re-enter the medium, escaping in a similar fashion to the first condition.
    \item The photon exits the medium at the outer boundary of the torus.
\end{enumerate}

The first escape condition is triggered when we calculate the initial angle of the photon trajectory to be less than the opening angle of the torus. Given a 3-dimensional direction vector \textbf{v} = ($v_x$, $v_y$, $v_z$), this condition is

\begin{equation}
    \theta_0 = \left| \text{arctan}\left( \frac{\sqrt{v_x^2 + v_y^2}}{v_z} \right) \right| < \theta_{\text{OA}}
\end{equation}

The second escape condition depends on both the photon position \textbf{r} = ($x$, $y$, $z$) as well as its direction \textbf{v}. The top (and bottom) of the hollow cut-out region is defined by a circle of radius $R_{\text{c}} = R_{\text{out}}\text{sin} \, \theta_\text{OA}$. This circle is a vertical distance $z_{\text{c}} = R_{\text{out}}\text{cos} \, \theta_\text{OA}$ above (and below) the origin. When a photon exits the inner wall of the medium, we first calculate the number of steps it must take to reach this $z_{\text{c}}$, assuming it continues in a straight-line path: 

\begin{equation}
    N_{\text{step}} = \frac{\text{sgn}(v_z) \times z_{\text{c}} - z}{v_z}
\end{equation}

\noindent where $\text{sgn}(v_z)$ indicates that we must choose the top boundary for photons traveling upward and the bottom boundary for those traveling downward. Then, we find the $x$ and $y$ coordinates after $N_{\text{step}}$ steps:

\begin{subequations}
\begin{equation}
    x_f = x + v_x N_{\text{step}}
\end{equation}
\begin{equation}
    y_f = y + v_y N_{\text{step}}
\end{equation}
\end{subequations}

\noindent Finally, if $x_f^2 + y_f^2 \leq R_{\text{c}}^2$, the photon will escape as it will travel through the open circle and out of the torus.

The third and last escape condition is simple. Let $|\textbf{r}| = x^2 + y^2 + z^2$. If $|\textbf{r}| > R_{\text{out}}$, the photon has exited the medium and will travel to infinity.

\end{document}